\documentclass{article}
\usepackage{graphicx} 
\usepackage{xcolor}
\usepackage{amsmath}
\usepackage[normalem]{ulem}
\usepackage{cite} 
\usepackage{placeins}
\usepackage{multibib}
\usepackage[pdfborder={0 0 1}]{hyperref}

\newcites{SM}{Supplementary References}

\title{Spontaneous epicuticular charging affects droplet dynamics on living leaves}
\author{Mihir Durve$^{1,\dagger, *}$, Serena Armiento $^{2,\dagger}$, Benham Kamare$^{3}$,\\ Sauro Succi$^{1,4}$, Barbara Mazzolai$^{2}$, Fabian Meder$^{3,*}$}

\date{\small
\textsuperscript{1} Center for Life Nano, $\&$ Neuro-Science, Fondazione Istituto Italiano di Technologia (IIT), 00161 Rome, Italy
\\
\textsuperscript{2} Bioinspired Soft Robotics Laboratory, Fondazione Istituto Italiano di Tecnologia, 16163 Genova, Italy
\\
\textsuperscript{3} Surface Phenomena and Integrated Systems, The BioRobotics Institute, Sant'Anna School of Advanced Studies, Pisa, 56216 Italy
\\
\textsuperscript{4} Department of Physics, Harvard University, Cambridge, MA 02138, United States 
\\
\textsuperscript{$\dagger$} these authors contributed equally
\\
\textsuperscript{$*$} corresponding authors: mihir.durve@iit.it, \\ fabian.meder@santannapisa.it}

\begin{document}
\maketitle

\section{Abstract}
How water droplets move and slide on leaves influences plant ecophysiological and abiotic interactions \cite{lenz_ecological_2022, Eglinton_Cuticle_1967, Shepherd_StressWax_2006, schefus_climatic_2005}, as well as the design of advanced bio-inspired wetting materials \cite{koch_multifunctional_2009, kreder_anti-icing_2016, Guittard_BioinspWetting_2025}. Despite cross-disciplinary relevance, current descriptions of the in situ dynamics of droplets on living leaves focus almost exclusively on surface structure and chemistry, treating the leaf as a static, electrically neutral substrate \cite{lenz_ecological_2022}. Here, three decades after the mechanistic discovery of the Lotus effect\cite{barthlott_purity_1997}, we show that a yet "hidden" force due to instantaneous electrical phenomena affect the dynamic droplet motion on living leaves. Using high-speed motion tracking and precision charge measurements, we show that droplets sliding on the pristine epicuticular wax layer on superhydrophobic \textit{Colocasia esculenta} leaves strongly charge affecting its dynamics, previously observed only on synthetic (highly electronegative fluorinated) surfaces. Droplets accumulate charges of $Q_{\mathrm{p, D1}} = -0.02 \text{ to} -0.15 \,\text{nC}$ per $30 \,\mu\text{L}$ droplet on pristine leaves. However, we specifically demonstrate the crucial role of the epicuticular wax layer plasticity: by a structural modification that decreases its roughness amplitude, the same leaves gain an impressive 30-40 fold enhancement in charge transfer (reaching $Q_{\mathrm{t, D1}} = -2.8 \text{ to} -5.2 \,\text{nC}$) slowing the droplet by half due to an estimated electrostatic force of $\approx 11$ $\mu$N dominating the resistive forces. The charge accumulation is surface-history-dependent and charge quantities per droplet are surprisingly similar or even exceeding those recently reported from artificial surfaces \cite{Li_2022, stetten_slide_2019, zhou_deposition_2025, hinduja_slide_2024, Xu_TriboelectricWetting_2022}. Our findings prove that electrostatic charging is a fundamental component of droplet-leaf interactions, opening new research directions from charge-affected leaf ecology to sustainable materials for droplet-based energy harvesting by tuning surface treatments and, moreover, for reducing soil contamination by improving droplet residing times on the leaf in agricultural spraying.

\section{Introduction}

The outer surface of plant leaves forms the world’s largest biological-atmospheric interface. The processes occurring at this surface are not only crucial for many lifeforms on Earth but also serve as an essential inspiration for innovations in surface engineering\cite{koch_multifunctional_2009, Nishimoto_BioinspiredSelfCleaning_2012, Guittard_BioinspWetting_2025, kreder_anti-icing_2016}. Among the multiple dynamic reactions occurring on this interface, fluid-structure interactions, like water-leaf interactions can have essential implications on plant survival and adaptation to stress and climate change: sticking of droplets that act as vehicles for pathogens, or persistent water films that occlude the surface, impeding gas exchange and light capture essential for photosynthesis, among many other, affect the organism's survival \cite{Riederer_2006,koch_multifunctional_2009,yeats_formation_2013,lenz_ecological_2022, Eglinton_Cuticle_1967, Shepherd_StressWax_2006, schefus_climatic_2005}. To counterbalance these threats, plants have evolved to precisely engineer their outermost surface that dominates the external fluid interactions including the epicuticular wax layer, demonstrating a remarkable phenotypic plasticity in response to their environment. Assessing the dynamic structure-fluid interactions such as water droplet contact line motion on intrinsically complex physical architectures of living plant leaves remains still a significant challenge\cite{lenz_ecological_2022, Snoeijer_MovingContactLines_2013, Tadmore_OpenProblems_2021}. This gap persists even though these properties have been successfully mimicked in artificial, lotus-, salvinia-, or nepenthes-like materials \cite{wong_pitcher_2011, koch_multifunctional_2009, kim_salvinia_2022, kreder_anti-icing_2016}. Consequently, dynamic events during water-surface interactions, such as spontaneous charging, are not explored on living, biological surfaces despite that they crucially affect artificial surfaces \cite{yatsuzuka_electrification_1994, Li_2022, lin_contact_2022, dratschow_liquid_2025}. Moreover, the behavior of droplet-leaf interactions is still frequently simplified to static investigations \cite{Dunkerley_Static_2023, lenz_ecological_2022}. Quantitative experimental data on structure-fluid interactions especially on complex surfaces in their in situ state are required to further support also the excellent progress recently made in computational prediction of soft flowing matter \cite{Tiribocchi_Boltzmann_2025} including the analysis of interactions of complex biological systems with their surroundings \cite{Falcucci_extreme_2021}.

Here, we developed experimental and analytical tools to quantitatively track the dynamic behavior of droplets on living leaves as function of the plasticity of the epicuticular wax layer and use them to highlight for the first time a fundamental effect of charge-induced droplet slow-down on the living leaf surface that is driven and affected by the epicuticular wax layer plasticity. In particular, the contribution of an electrostatic force during droplet sliding has, until now, only been quantified for hydrophobized, engineered materials (typically fluorinated and micro-nano-structured), given its known effects on wetting problems in industrial processes and potential for applications like energy harvesting \cite{yatsuzuka_electrification_1994, lin_contact_2022}. Especially on fluorinated polymers like polytetrafluorethylene or oxide surfaces treated with a perfluorinated silanes, high voltages have been measured in sliding droplets due to spontaneous contact electrification of the surface \cite{li_sparking_2023, bista_high_2023}, and the resulting surface potentials can generate electrostatic forces that drastically affect the sliding speed \cite{Li_2022} and even overcome gravity \cite{miljkovic_electrostatic_2013,sun_surface_2019, Xu_TriboelectricWetting_2022}.
The underlying liquid-solid contact charging causing the electric fields is strongly material-dependent \cite{burgo_where_2016, dratschow_liquid_2025, lin_contact_2022}, but even on homogeneous artificial materials, the mechanism is still not well understood \cite{dratschow_liquid_2025, lacks_long-standing_2019}, especially effects of surface history \cite{Sobarzo_2025}, in addition to being studied for more than a century \cite{Thomson_ElectricityDrops_1894}. The dynamic charging of biotic surfaces is significantly complex and less investigated. However, recent studies show that plant-based materials like waxes and resins accumulate also cause significant liquid-solid contact charging \cite{Kamare_wax_2025, armiento_liquid-solid_2022}, on the other hand lignin-rich wood showed a strongly reduced tendency to contact electrification \cite{mertcan_ozel_why_2020}. Indeed, droplets hitting and spreading on leaves charge the leaf surface through liquid-solid contact electrification of the epicuticular waxes \cite{armiento_liquid-solid_2022}. Quantification of in situ electricifation of leaf surfaces is crucial as electrostatic forces have important implications on above-mentioned wetting-related phenomena including natural processes, such as pathogen dispersal on leaves \cite{park_dynamics_2020,mukherjee_synergistic_2021, kim_vortex-induced_2019} and also technological applications, such as increasing the efficiency of agricultural spraying \cite{jayaprakash_enhancing_2025} to even the improvement of biohybrid energy harvesting \cite{armiento_device_2023, wu_fully_2020, choi_spontaneous_2017}.
To tackle these problems, we combined two methods: automated, object-detection-based high-speed video tracking to analyze droplet dynamics \cite{durve_2022, Durve2023, Durve_2024_POF}, and precision current measurements to quantify charge accumulation from water droplets sliding down living \textit{Colocasia esculenta} (Taro) leaves. We chose this species as a model for two reasons: its large leaves provide a sufficient, reproducible sliding path, and its lotus-like superhydrophobicity has been shown to affect charge formation \cite{armiento_liquid-solid_2022}. 
We recorded a large dataset of over $300$ high-speed motion tracking videos of individual droplets, analyzed them with machine learning-assisted automatization as mentioned above and compared the results with precision charge measurements of about $25000$ droplets on 12 individual leaves from different plants. Our electrical acquisition is analogous to that used by Li et al. \cite{Li_2022} to track the charging of sliding droplets on artificial materials but we adapted the setup to measurements on living plants. We observed that, especially the first droplets sliding on a fresh leaf surface, are slower and a velocity increase with the number of sequential droplets sliding down the same path, simultaneously with a decrease in droplet charge accumulation. Importantly, to further confirm the role of epicuticular waxes on charging and droplet dynamics, we modified the plasticity of the epicuticular wax layer by a thermal smoothening of the surface roughness amplitude, leading to strongly enhanced droplet-surface  charging. Surprisingly, the charge quantities measured per single droplet occasionally even exceeded values reported on fluorinated artificial substrates \cite{Li_2022, bista_high_2023}. The increased charging in turn reduced droplet sliding speed on the leaves significantly. Moreover, our study shows that automated object tracking analysis can give additional information, such as droplet width, on dynamic variations in the interaction of droplets with surface, analyzing bounding box parameters to derive information on surface interactions. 

Overall, our results clearly show for the first time that electrostatic interactions arising from in situ electrification of the epicuticular wax layer on the surface of living leaves play a fundamental role in droplet-leaf dynamics. As one of the largest and most crucial biological interfaces for all life on Earth, this opens up further research questions such as if the wax layer on some species specifically evolved to exploit electrification to control external water interactions. How  heat, erosion or photo damage of the wax layer due to climate change affect the charging and fluid dynamics of the leaf surface and subsequent ecophysiological interactions like stress response and climate monitoring\cite{Shepherd_StressWax_2006, schefus_climatic_2005}. Moreover, how can charging phenomena on leaves be exploited in improving droplet retention on the leaf in precision agricultural spraying, and how leaf surfaces or bioinspired materials that are more sustainable as fluorinated materials could be used in droplet energy harvesting.

\pagebreak

\section{Methods}

\textbf{Plants:} \textit{Colocasia esculenta} were grown from mature, dormant corms (diameter: 5 - 8 cm) in a controlled greenhouse environment (MG322, Monti Co., Italy, maintained at 27 °C, 50 \% relative humidity, 16 hours photoperiod). Corms were planted in 1 L plastic pots filled with a standard potting mix and soil was kept moist by watering every 1-3 days. The plants were grown until leaves reached a lamina length of about 25-35 cm from petiole junction to apex to obtain leaves that are large enough to provide a flat, homogeneous sliding surface for a length of at least 4 cm that is not significantly interrupted by veins. Each plant typically produced 1-3 suitable leaves in a growth period of 3-6 weeks. For the experiments, leaves were excised by cleanly cutting at the base of the petiole, as close as possible to the corm or soil interface and transferred to a water-filled tube (15 mL Falcon tube) containing deionized water and submerged to maintain turgor pressure and physiological viability. The experiments were typically conducted within 1 - 4 hours post-harvest, no visible wilting or loss of rigidity was observed within this time window.

\textbf{Modification of the epicuticular waxes:} The epicuticular wax layer was modified directly on the living leaf by a physical, thermally-induced structural smoothening as previously described \cite{armiento_liquid-solid_2022}. Wax crystals were heated on the living leaf by exposure to hot air stream of about 180°C  for $\approx 3s$ (applied in a distance of 10 cm from the leaf surface) sufficient to change the nanoscale wax crystal structure and smoothening the surface (bulk melting point of \textit{Colocasia esculenta} waxes $\approx 65^\circ C$) without major damage to the tissue and comparable tissue properties to pristine leaves for the duration of the experiments\cite{armiento_liquid-solid_2022}.  

\begin{figure}[ht]
  \centering
  \includegraphics[width=\textwidth]{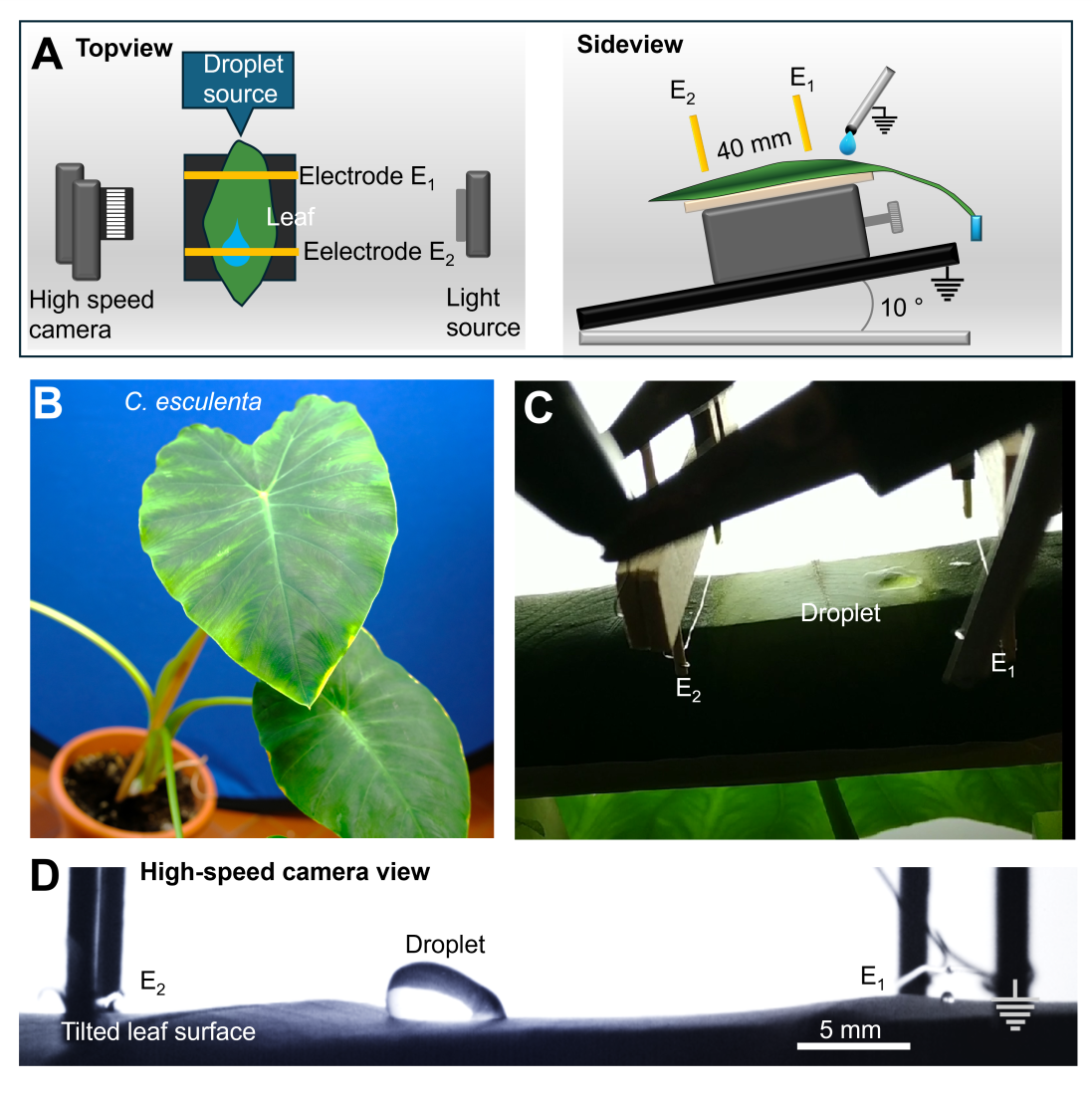}
  \caption{Experimental setup and tracking of droplet sliding on pristine \textit{C. esculenta} leaves. a) Schematic of the test setup in which a fresh leaf was fixed on a tilted antistatic sample holder enabling a homogenous sliding path of 40 mm, a droplet source was installed about 10 mm above the leaf and two electrodes ($E_1$ for neutralizing droplets before the slide and $E_2$ for measuring accumulated charges. A high speed camera operating at 1200 fps was used for tracking droplet dynamics. b) Photograph of a \textit{C. esculenta}, the superhydrophobic leaves here have a length of 25-30 cm providing a sufficient area for the experiment. c) Photograph of a typical experiment: the leaf installed in the acquisition system, showing sliding path, droplet and electrodes. d) Sideview image recorded by the high-speed camera. Videos 1 and video 2 show the sliding motion and give an example of the tracking using an bounding box.}
  \label{ref_fig1}
\end{figure}

\textbf{Setup for leaf electrification analysis and video recording:} During experiments, the leaves were mounted on a  tiltable optical breadboard by gently wrapping the lamina around a wooden cylinder (diameter: 5 cm) flattened at its upper surface (see Figure \ref{ref_fig1}) while keeping the leaf hydrated as described above. When necessary, the leaf was lightly secured using paper tape to maintain position without inducing mechanical damage. The holder was specifically designed to provide a reproducible, flat sliding path for droplets and wood was selected as the support material due to its low tendency to accumulate static charge, thereby reducing interference during electrical measurements \cite{mertcan_ozel_why_2020}. The platform on which the high-speed camera (Miro C110, Phantom High Speed, USA) and a backlight were mounted, was kept at an inclination of 40°. Monochrome videos were recorded at 1,200 frames per second (fps). The setup was placed in a Faraday cage. Droplets of a fixed volume ($30\,\mu\text{L}$) were precisely dispensed via a $50\,\text{mL}$ syringe coupled to a syringe pump (Linari, Italy). The fluid stream was passed through a grounded $4\,\text{mm}$-diameter metal tube, positioned $1-2\,\text{cm}$ above the first electrode, to actively neutralize any residual electrostatic charges acquired during transit through the syringe and silicone delivery system. To establish robust statistics, experiments were systematically conducted and reproduced across 12 individual leaves from distinct plants, as detailed in the results section. For each specific leaf and condition (e.g., pristine or treated), currents were measured from a minimum of $900$ sequential sliding droplets, complemented by the analysis of over $300$ high-speed videos capturing single droplet dynamics. The presented data are representative, and all derived conclusions are supported by the specified statistical analysis.

\textbf{Electrical measurements:} Droplet discharge currents have been recorded after droplets slide for 40 mm using a 0.3 mm platinum electrode (pt00-wr-000128, GoodFellow, UK) using a high input impedance electrometer (6517B, Keithley, USA) combined with an oscilloscope (MSO7014A, Agilent Technologies, USA). Two platinum electrodes, electrode 1 at the beginning of the 4 cm slide and electrode 2 at the end, were installed floating at a distance of about 2 mm above the leaf surface. Electrode 1 discharged the droplet through a 1 $M\Omega$ oscilloscope probe, electrode 2 was used to measure the droplet charge accumulated during sliding for 40 mm by integrating over the discharge current spike recorded with the electrometer.

\textbf{Droplet tracking and video analysis:} Droplet identification and tracking in sequential frames were performed using DropTrack, a specialized software tool \cite{durve_2022,Durve2023, Durve_2024_POF}. DropTrack leverages state-of-the-art YOLO (You Only Look Once) \cite{yolo_original} and DeepSORT algorithm \cite{deepsort_original}, optimized for accurate droplet detection and robust tracking. The output from DropTrack includes bounding box coordinates that tightly enclose each droplet and a unique identifier assigned to each. This information facilitates the extraction of velocimetric data from the image sequences. For this study, the YOLOv11l model \cite{yolo11_ultralytics} was trained on approximately 1121 manually annotated images, supplementing a base training dataset from previous experiments \cite{droplet_data}. See Supporting Information Figure \ref{ref_figs1} for the details on YOLO model training. Representative tracking videos are available in the supplementary material (Video 2, pristine and 4, treated leaf). The high-speed videography using a constant distance between the leaf and the camera ensured consistent imaging conditions throughout all tests. The droplet rolling distance was standardized at 40 mm, defined by the separation of the discharging electrodes. This fixed configuration allowed the measurement of various physical observables without requiring additional external hardware. Specifically, we measured three key parameters: the total distance traveled by a droplet at a given time, the instantaneous speed of the droplet’s projected centroid, and the droplet width as it slid across the leaf surface. The reported distance and time measurements correspond to the center of mass of the droplet, which was approximated as the center of the bounding box determined by the tracking algorithm.

\section{Results}

\begin{figure}[ht]
  \centering
  \includegraphics[width=\textwidth]{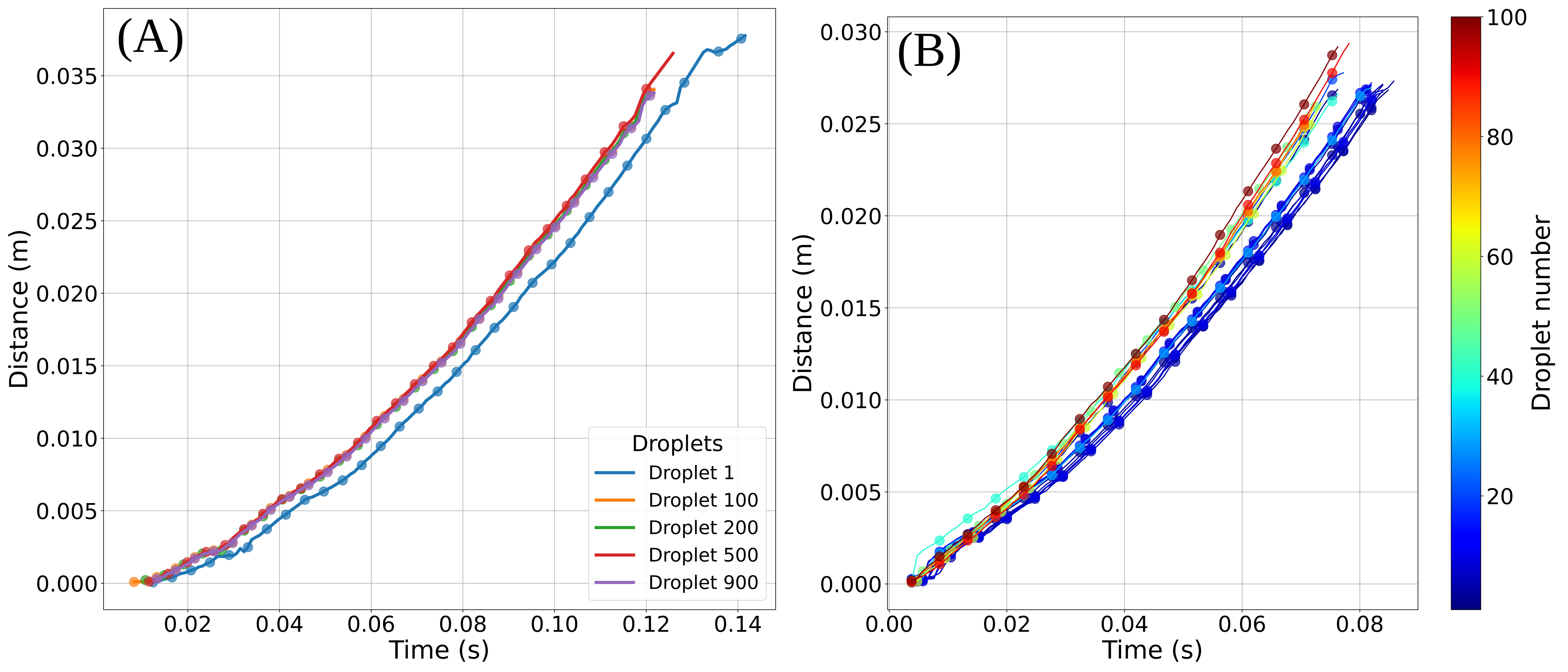}
  \caption{Droplet motion tracking on superhydrophobic pristine leaves. a) Total distance traveled by droplets at a given time on the same leaf and sliding zone. b) Higher detail analysis of droplet dynamics and total distance traveled by droplets D1-D100 showing a spatiotemporal evolution of droplet displacement with droplet number and slowdown of the first droplet.}
  \label{ref_fig2}
\end{figure}

\subsection*{Dynamics of droplet sliding on pristine leaves}
Our tests initially focused on high-resolution motion tracking of droplets sliding on superhydrophobic leaves. Figure \ref{ref_fig1}a shows the experimental setup in which a \textit{C. esculenta} (Figure \ref{ref_fig1}b) leaf was fixed at a 40° tilt angle providing a homogeneous 40 mm sliding path for the tests (Figure \ref{ref_fig1}c).
A single droplet (30 $\mu$L) was generated at a height of approximately 1 cm from the leaf surface, landed on the tilted leaf where it started to slide. Subsequent droplets generated at a frequency of $\approx$0.8 Hz  typically took the same sliding path (sliding of about 900 droplets was tested on the same leaf). The droplet dynamics were analyzed by high speed video recordings (frame snapshot in Figure \ref{ref_fig1}d, recorded at 1200 fps, Video 1 and Video 2 show example videos of a droplet sliding on a pristine leaf at 40° inclination with tracking, respectively) and we extracted the velocimetric data from the image sequences using computer vision based DropTrack software (further detailed in Supporting Information Figure \ref{ref_figs1}) to predict bounding boxes to localize and identify droplets and to extract parameters like droplet speed, acceleration, and droplet width that was applied to $>300$ high-speed videos. Figure \ref{ref_fig2}A shows the traveled distance of the droplet as function of sliding time for droplet 1 to 900 (D1 to D900). As expected, droplets generally accelerated consistently during their slide. However, the first droplet on the path was typically slowest, taking up to $8\%$ longer to travel a 20 mm distance than later droplets in the path history. The travel times for D100 to D900 on the same path no longer varied significantly. This indicates that the dynamics depend on the surface history and essential variations in droplet dynamics occur primarily within the first 100 droplets. Thus, we tracked the first 100 droplets in higher detail as given in Figure \ref{ref_fig2}B. Our results, as shown in panels A and B, are derived from distinct experimental replicates, each utilizing a new leaf under otherwise identical conditions. Nonetheless, we consistently observed the initial droplet's deceleration across all experimental replicates (further tracking results on nine individual leaves are given in Supporting Figure  \ref{ref_figs2}A). The results confirm a stepwise reduction of the time needed to slide the same distance for sequential droplets, with the first droplets being the slowest by a factor of 1.08 compared to the 100th droplet and observing increasing velocity with droplet number. Next, we aimed to gain deeper insight into the role of the epicuticular wax layer on the same leaf and sliding path.

\begin{figure}[ht]
  \centering
  \includegraphics[width=\textwidth]{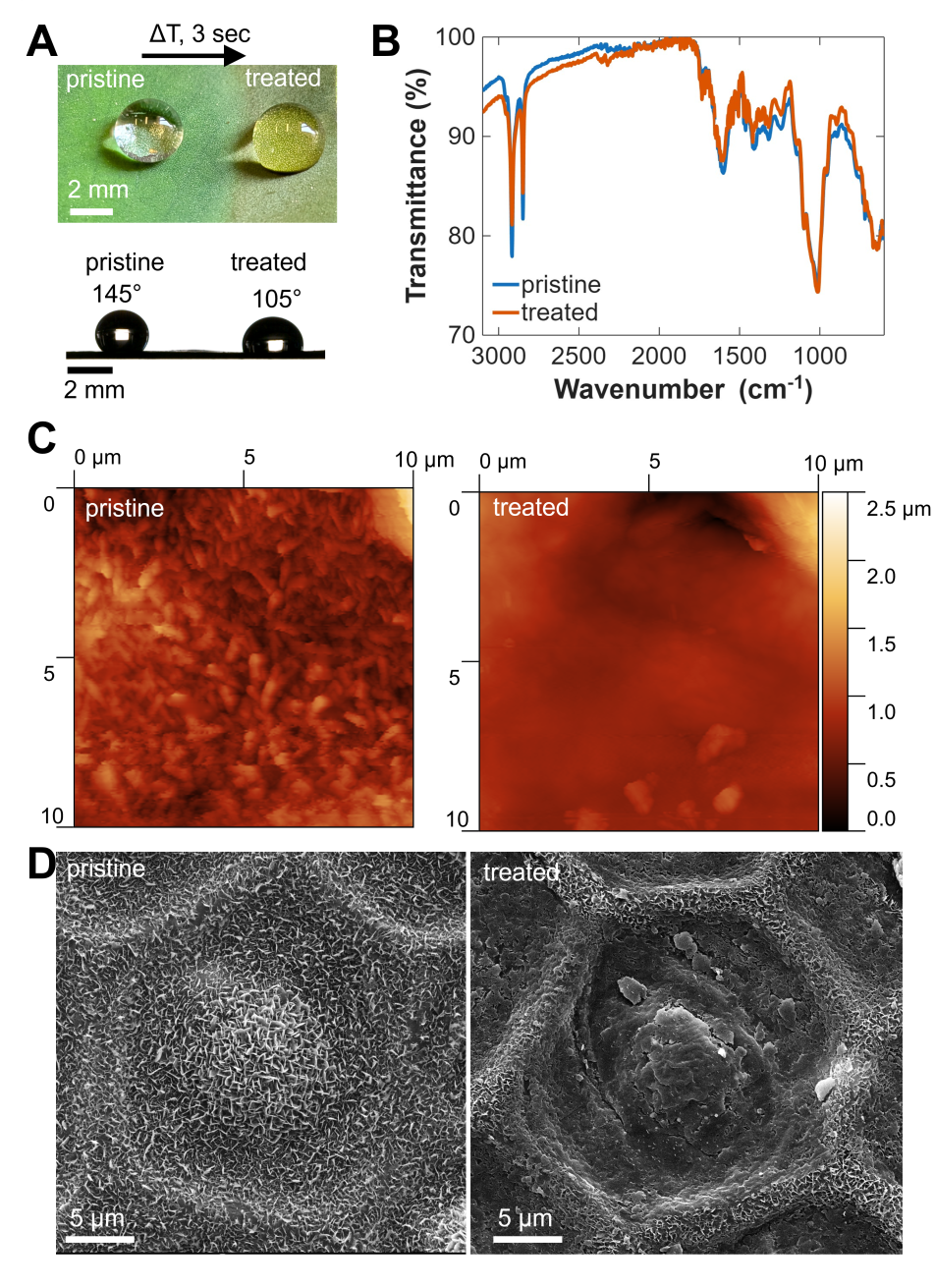}
  \caption{Modification of the epicuticular wax layer plasticity and effect on droplet dynamics. A) Images of static contact angle of a water droplet sitting on a pristine leaf and the same leaf after surface treatment showing the reduction in the contact angle, $\Delta \theta = 40^{\circ}$, due to a structural smoothening of the surface waxes. B) FTIR measurements on the leaves before and after treatment confirming that the overall surface chemistry did not change significantly suggesting that mostly the morphology of the epicuticular wax layer was affected. C) AFM topography images of the same area of the leaf surface before (pristine) and after modification (treatment) showing a distinct surface smoothing and decrease in roughness amplitude (Extended data Figure \ref{ref_figs3} shows roughness profiles). D) SEM images of pristine and treated leaves showing the variation of the nanostructured wax layer towards a smoother configuration.}
  \label{ref_fig3}
\end{figure}

\subsection*{Dynamics of droplet sliding on leaves with modified epicuticular wax plasticity}
\begin{figure}[ht]
  \centering
  \includegraphics[width=\textwidth]{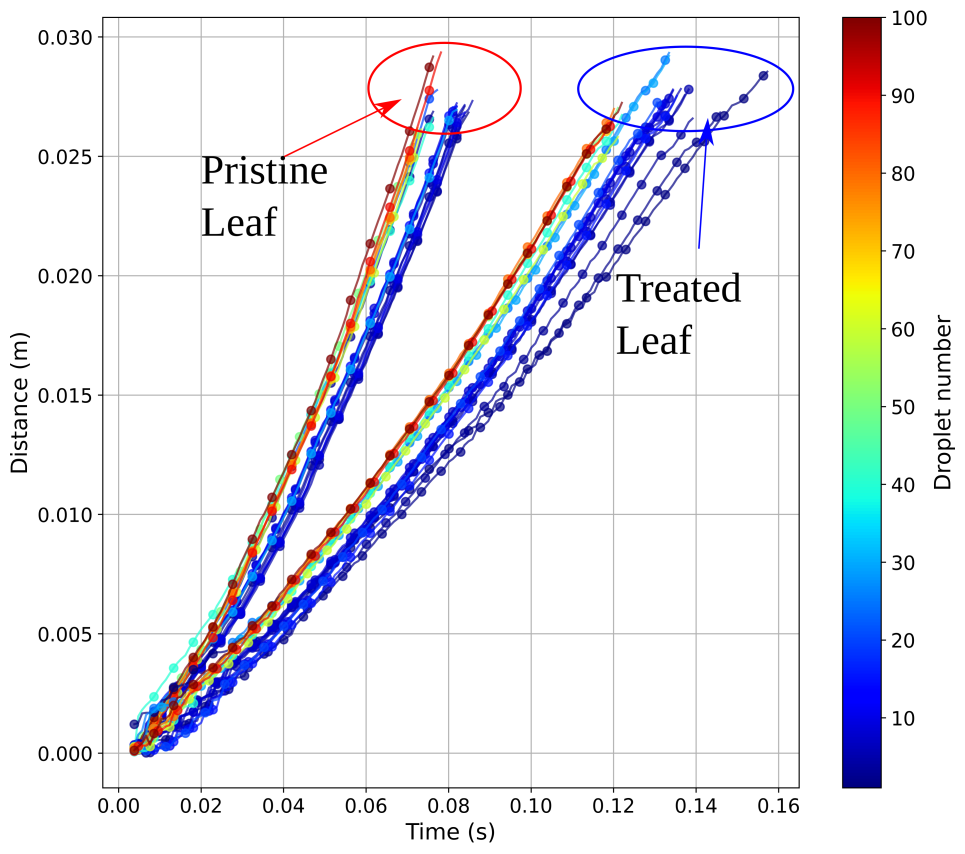}
  \caption{Droplet travel distance as a function of surface treatment. This figure compares the motion of 100 droplets on a pristine leaf (highlighted by a red circle) with a treated leaf (shown in blue circle) showing I) a general droplet slowdown after surface treatment in addition to II) a transient behavior in the slowdown which reduces with droplet number. The zones in which the droplets slide are exactly the same before and after treatment so that the effect of the epicuticular wax layer modification can directly be observed.}
  \label{ref_fig4}
\end{figure}

The epicuticular waxes on the surface of \textit{Colocasia} leaves expectedly govern droplet motion by influencing contact-angle hysteresis, droplet footprint, and retention forces, while also reducing contact-line friction and determining roll-off and pinning behavior \cite{neinhuis_characterization_1997,koch_multifunctional_2009, lenz_ecological_2022}. To further investigate their role in the dynamic liquid-surface interaction, we modified the wax plasticity using a gentle thermal treatment that smoothens the uppermost nanostructured wax crystal layer (previously introduced by us \cite{armiento_liquid-solid_2022}). Essentially, the treatment allows to measure the effect of the wax plasticity before and after modification on the same leaf and exactly the same position. This decreases the surface roughness by smoothening the epicuticular wax crystals due to thermal fusion into a more homogeneous layer resulting in a surface that remains hydrophobic but with a lower static contact angle of approximately 105° ($\Delta \theta = 40^{\circ}$ compared to the pristine leaf surface), as shown in Figure \ref{ref_fig3}a. Fourier transform infrared spectroscopy (FTIR) measurements in Figure \ref{ref_fig3}b suggest that the overall surface chemistry has been maintained during the treatment as no significant changes in the absorption bands are seen, whereas the AFM and SEM imaging given in Figure \ref{ref_fig3}c and \ref{ref_fig3}d, before and after treatment confirm a clear morphological variation in the leaf microstructure due to smoothing of the wax layer. AFM measurements were taken at the same position on the leaf before and immediately after treatment.
The effect on droplet dynamics are as follows. The modification caused a significant droplet slowdown (Video 3 shows a comparison of droplet sliding on pristine and treated leaf surface of the same leaf, Video 4 gives another example of droplet sliding on the treated leaf surface and the tracking with a bounding box). As anticipated from a lower hydrophobicity, the droplets slide slower than on the pristine superhydrophobic surface, and the slope of the droplet speed over time, decreased by about $60\%$ compared to the pristine surfaces (see Fig. \ref{ref_fig4}). However, again, we observe a significant transient increase in the sliding velocity over the first 100 droplets for sequential droplets on the same leaf and systematic differences in the distance traveled at a fixed sliding time across successive droplets (see Fig.\ref{ref_fig4}). By  t=$100\,\mu\text{s}$, D100 has traveled $\sim5$ mm farther than D1, sliding off the leaf more efficiently. In line with this, D100 reached a sliding distance of 25 mm $25\%$ faster than D1. From D100 on to D900, no significant further increase in droplet speed has been observed (Extended Data Figure \ref{ref_figs2}B).

The reduction in droplet speed can be conceptualized by mapping the dynamics onto a classical physical system: a rigid sphere sliding down a frictionless inclined plane (Extended Data Figure \ref{ref_figs4}). For the pristine leaf, the motion of the first droplet follows the trajectory of an object on a $4^\circ$ effective frictionless incline (given an initial speed of $0.29 \pm 0.06 \text{ m/s}$ imposed by the pumping system). In other words, the resistive forces at the water-leaf interface reduce the apparent inclination from $40^\circ$ to $4^\circ$, quantifying the degree of kinetic energy dissipation achieved by the surface treatment. Conversely, the dynamics observed on the treated surface correspond to an object moving along a $2^\circ$ frictionless inclined plane with an initial speed of $0.15 \text{ m/s}$. The variation of resistive forces as function of droplet sequence and surface history are analyzed later in detail.

\begin{figure}[ht]
  \centering
  \includegraphics[width=\textwidth]{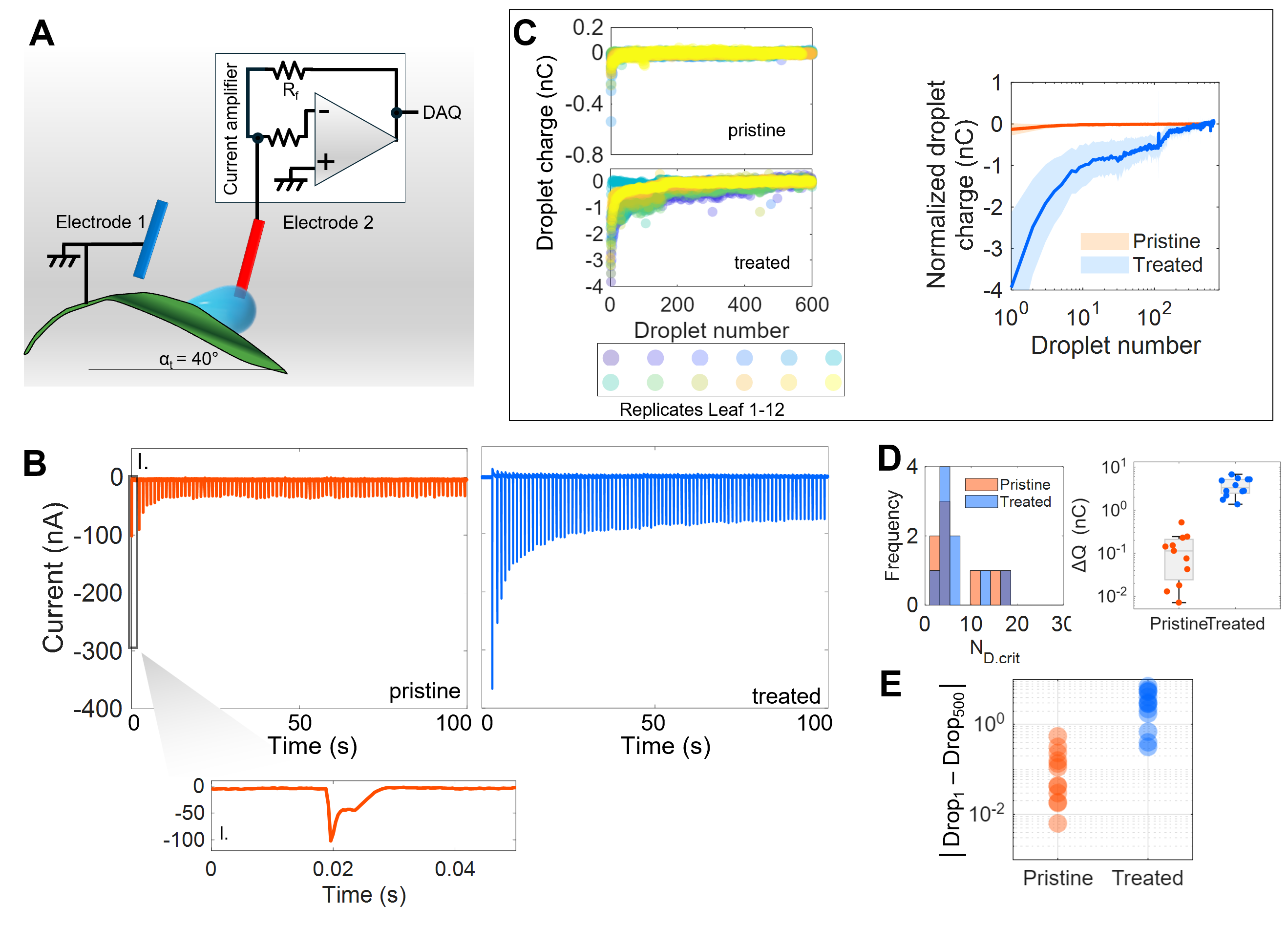}
  \caption{Spontaneous charging of droplets sliding on pristine and modified \textit{C. esculenta} leaves. a) Schematic of the measurement circuit used to measure droplet sliding on the leaf tilted at 40° on a path of 40 mm between electrode 1 (grounded to initially neutralize the droplet) and electrode 2 used to acquire the discharge current after the slide. b) Typical currents measured at electrode 2 for the pristine (left panel) and modified leaf (right panel) when exposed to multiple droplets sequentially sliding on the tilted leaf. Each droplet crossing electrode 2 produces a single, well-distinguishable current peak (zoom in of first droplet`s peak in lower panel). A clear decrease in amplitude can be seen with peak and droplet number for both, pristine and treated leaf surface; modification of the epicuticular waxes leads to higher current amplitudes. c) Detailed analysis of charges accumulated in the droplets during the slide as function of droplet number of leaves from different plants (n = 12); upper left panel pristine, lower left panel treated. The right panel shows mean (colored curve) and standard deviation (colored area) of charge accumulation as function of droplet number comparing pristine (orange) and treated (blue) leaves. The results indicate the decrease of droplet charge with droplet number sliding on the same path. d) Histogram of $N_D,crit$ distribution and $\Delta Q$ for pristine and treated leaves, respectively obtained by fitting the charges measured in droplet sequences on individual leaves (n = 12), indicating that the variation in the wax layer increases the charging capacity while concurrently slowing the saturation kinetics, as reflected by the increase in the characteristic droplet number. e) Point clouds showing the absolute difference in the droplet charge between drop D1 and drop D500 on different pristine and treated leaves (n = 15).}
  \label{ref_fig5}
\end{figure}

\subsection*{Slide-induced droplet charging and effect on droplet motion}
To further analyze what causes the variation in droplet dynamics on the leaf surface, we tracked the charge accumulation in the sliding droplets using a two-electrode setup (Figure \ref{ref_fig1}a–\ref{ref_fig1}b) and the measurement circuit in Figure \ref{ref_fig5}a. Before sliding, the droplet was discharged twice: first at the grounded outlet of the dispensing tube connected to the droplet generator and again by electrode 1 (E1) at the start of the slide. Electrode 2 (E2) was used to measure discharge current spikes after sliding for 40 mm, enabling high-resolution and in-situ current measurements on the living leaf with simultaneous video recordings.
Figure \ref{ref_fig5}b shows typical current signals recorded for sequential droplets on pristine and treated leaves (left and right panels, respectively) for the first 80–100 droplets. The lower panel shows the current peak due to the first droplet on pristine leaves in high detail, revealing a peak duration of 9 ms and an initial sharp discharge peak. The negative current spikes suggest the accumulation of a negative charge in the droplets sliding on the leaves. This is likely due to the deposition of a positive charge through contact electrification of the leaf surface, as previously reported \cite{armiento_liquid-solid_2022}. The trend is similar on the pristine and treated leaves, but discharge current spikes are significantly higher on the treated surface (here $\approx$350 nA) compared to ($\approx$115 nA on the pristine leaves) considering the first droplet.
Figure \ref{ref_fig5}c shows the net charges ($\int I(t) dt$) transferred by droplets across $n=12$ leaves, with a 600-droplet sequence per leaf. The analysis tracks droplets on the same sliding zone for both pristine (upper left) and treated samples (lower left). The corresponding individual discharge current measurements on leaves from different plants are given in the Extended Data Figure \ref{ref_figs5} and \ref{ref_figs6} for pristine and treated leaves, respectively. 

Remarkably, the variation in the wax structure leads to a consistent enhancement throughout the entire droplet sequence. We observe an average 30 to 40-fold gain (observed in $n=12$ leaves), $G = (Q_{treated} / Q_{pristine})-1$, in both, the initial charging of the first droplets and the charge at saturation for droplet numbers $>100$ (see Extended Data Figure \ref{ref_figs7}). This indicates a significant increase in effective surface interaction sites that cause charging due to the variation in the wax layer plasticity after treatment. The quartiles of the measured charges of the first droplet across all pristine leaves lie in the range of $Q_{\mathrm{p, D1}} = -0.02 \text{ to} -0.15 \,\text{nC}$, whereas those for treated leaves significantly higher with $Q_{\mathrm{t, D1}} = -2.8 \text{ to} -5.2 \,\text{nC}$ per $30 \,\mu\text{L}$ droplet. The minimum and maximum range values mentioned above are given as the first quartile and the third quartile respectively, computed over the measurement of 12 individual leaves. To highlight the extremes, for D1 on pristine leaves, highest charging of -0.54 nC (lowest 0.014 nC) was detected on a single specimen (see Extended Data Figure \ref{ref_figs5}) while D1 on treated leaves highest droplet charge measured was -6.8 nC (lowest 1.4 nC, see Extended Data Figure \ref{ref_figs6}). Individual leaves showed gain factors $G$ of over 100 comparing the pristine and treated surface for the first droplet. 

Interestingly, the charges measured on the treated leaves are even higher than most droplet charges so far reported on fluorinated (and highly electronegative) artificial substrates. Extended Data Figure \ref{ref_figsX} compares literature data on artificial surfaces (mainly fluorinated glass surfaces) that have been obtained under comparable experimental conditions with our results confirming this trend. The data not only proves a significant difference caused by the wax layer variation but also again that the first droplets accumulate most charges across all tested conditions which is further demonstrated in the cumulative uncertainty graph in Figure \ref{ref_fig5}c, right panel, constructed from the mean and standard deviation of droplet charge measurements on the 12 individual leaves from different plants. 

Additionally, the charge saturation in a sequence of droplets sliding down the same path, indicates that available surface charging sites are being depleted. However, the effect is reversible: the charging behavior recovers when the same sliding path is tested after 60 minutes, indicating that the charge transfer sites on the wax layer at the droplet-leaf interface regenerate (see Extended Data Figure \ref{ref_figs8}).

The variation in the structure of the epicuticular wax layer significantly alters the electrification of the droplets compared to the pristine surface: 1) The first droplet yields the peak surface charge in both cases; however, this value is at least tenfold lower when measured on pristine leaves. 2) Charge saturation as a function of droplet number occurs significantly earlier on pristine leaves compared to the treated surfaces. 3) To further quantify the saturation kinetics shown in Figure~\ref{ref_fig5}d, we fitted the data to the equation $Q(N_D) - Q_0 = \Delta Q \left[ 1 - \exp\left(-N_D / N_{D, \mathrm{crit}}\right) \right]$ were $Q(N_D)$ is the instantaneous charge measured on the \textit{N}-th droplet, $N_D$ is the droplet number, $Q_0$ is the initial charge offset typically close to zero, $\Delta Q$ is the total capacity (sites) on the surface for charge accumulation, and $N_D,crit$ is the number of droplets required to reach ~63 \% of the total saturation amplitude $\Delta Q$. The analysis yields average fit parameters of $\Delta Q \approx 0.14 \pm 0.15 \mathrm{nC}$ and $N_{D,\mathrm{crit}} \approx 8 \pm 6$ for 12 pristine leaves, compared to $\Delta Q \approx 3.97 \pm 1.63\mathrm{nC}$ and $N_{D,\mathrm{crit}} \approx 13 \pm 20$ for 12 treated leaves (see Figure~\ref{ref_fig5}d). Although there is expected variation in individual test results for different leaves, the trend is clear and reveals that while the change in the wax layer plasticity drastically increases the effective charging sites (indicated by the order-of-magnitude rise in $\Delta Q$), it also lowers the saturation kinetics. The increase in $N_{D,\text{crit}}$ may suggest that the denser network of surface sites available for effective charge transfer requires a higher number of cumulative droplet interactions to reach saturation. This is likely due to the increased effective surface area accessible for charge transfer upon the transition from a superhydrophobic to a better wetting state.

\subsection*{Electrostatic contribution in dynamic resistive forces on living leaves}
\begin{figure}[ht]
  \centering
  \includegraphics[width=\textwidth]{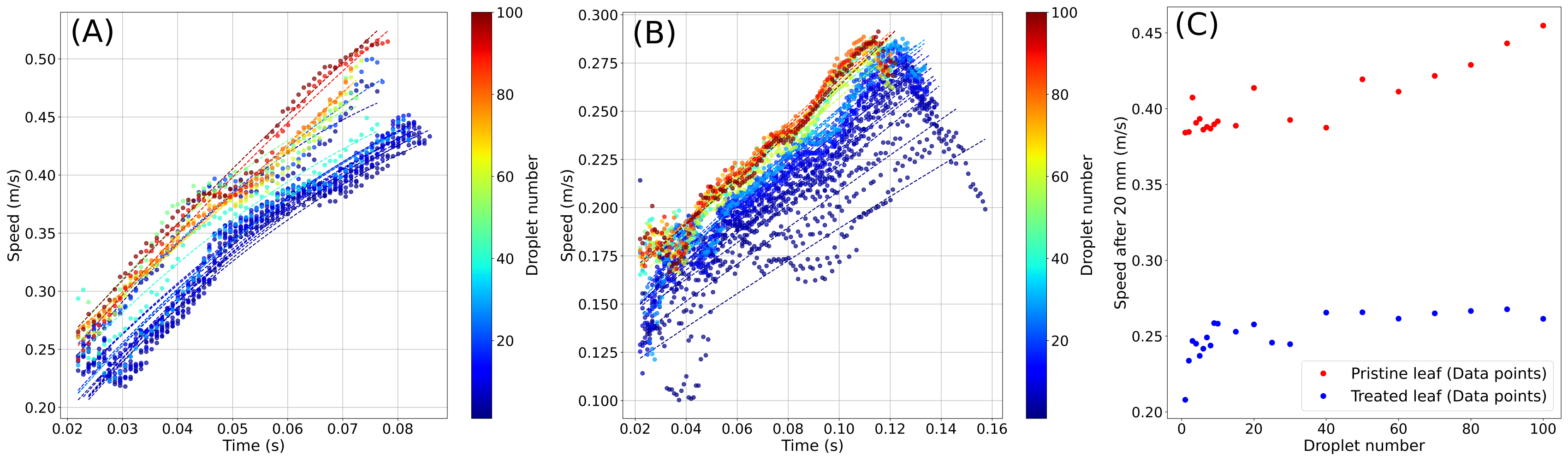}
  \caption{Instantaneous speed of droplets measured across the duration gap of 20 frames on a) pristine leaf, b) treated leaf as function of sliding time and droplet number. c) instantaneous speed of droplets after sliding a distance of 20 mm from the first electrode as function of droplet number clearly showing the effect of the surface treatment on the droplet transport dynamics and a surface history-dependent increase in droplet transport on the leaf surface. The dashed lines are 0th order best fit serving as guides to the eye to observe the trend of increasing speeds.}
  \label{ref_fig6}
\end{figure}

Importantly, the electrical measurements closely mirror the results of the droplet speed and traveled distances from video tracking. As shown in Fig. \ref{ref_fig6}A and Video 5, the first droplet on a pristine leaf is slowest and droplets speed up with larger droplet numbers, similar to what has been observed during the reduction in charge transfer with repeated droplet passages. Similarly, the droplets velocities are decreased after surface treatment (Droplets on pristine leaves exceed instantaneous speed of 0.42 m/s, whereas on treated leaves they remain below 0.29 m/s. The speed measurements are taken across 20 frames and the values mentioned above are measured after sliding 40 mm distance), as shown in Figure \ref{ref_fig6}B and Figure \ref{ref_fig6}C illustrates the speed difference at a fixed point located 20 mm from the first electrode (E1). The videos comparing drop 1 and 100 in Video 5 and Video 6 for pristine and treated leaves, respectively clearly visualize the effect.

To further quantify the observation of a connection of droplet motion and charging (Extended Data Figure \ref{ref_figs9}), we compared the droplet acceleration to the ideal frictionless limit ($a_{\text{ideal}} = g \text{sin}(\theta)$) to derive the instantaneous resistive force $F_{res}$ as a sum of multiple coexisting forces such as contact line friction, viscous dissipation, and possible electrostatic forces slowing down the droplets when sliding in sequence and on pristine and treated leaves using Newton's Second Law $F_{res} = m (g \sin\theta - a_{actual})$, details in the Supporting Information and in Extended Data Figure \ref{ref_figs9}. Analyzing a sequence of 100 droplets sliding on the pristine and subsequently the treated leaf, shows that the surface treatment increases the overall magnitude of $F_{res}$ doubling the initial interaction force from 83  $\mu$N to 164 $\mu$N. Second, the surface treatment and the consequential higher charging allow detecting a clearly measurable exponential decay of $F_{res}$ (see Extended Data Figure \ref{ref_figs9}). Interestingly, the mechanical decay constant derived from this model ($b \approx 0.18$) and the charge saturation rate ($1/N_{D,crit} \approx 0.15$) under the same conditions follow exactly the same trend with almost the same numerical value. This clearly suggests that the higher initial resistive force is governed and driven by surface charging of the wax layer on the leaf. Consequently, one can estimate the specific electrostatic contribution to $F_{res}$ on the treated leaves which is approximately 11 $\mu$N, a magnitude consistent with electrostatic resistive forces reported for artificial fluorinated surfaces under similar experimental conditions \cite{Li_2022}.

\subsection*{Temporal variation of bounding box width}
\begin{figure}[ht]
  \centering
  \includegraphics[width=\textwidth]{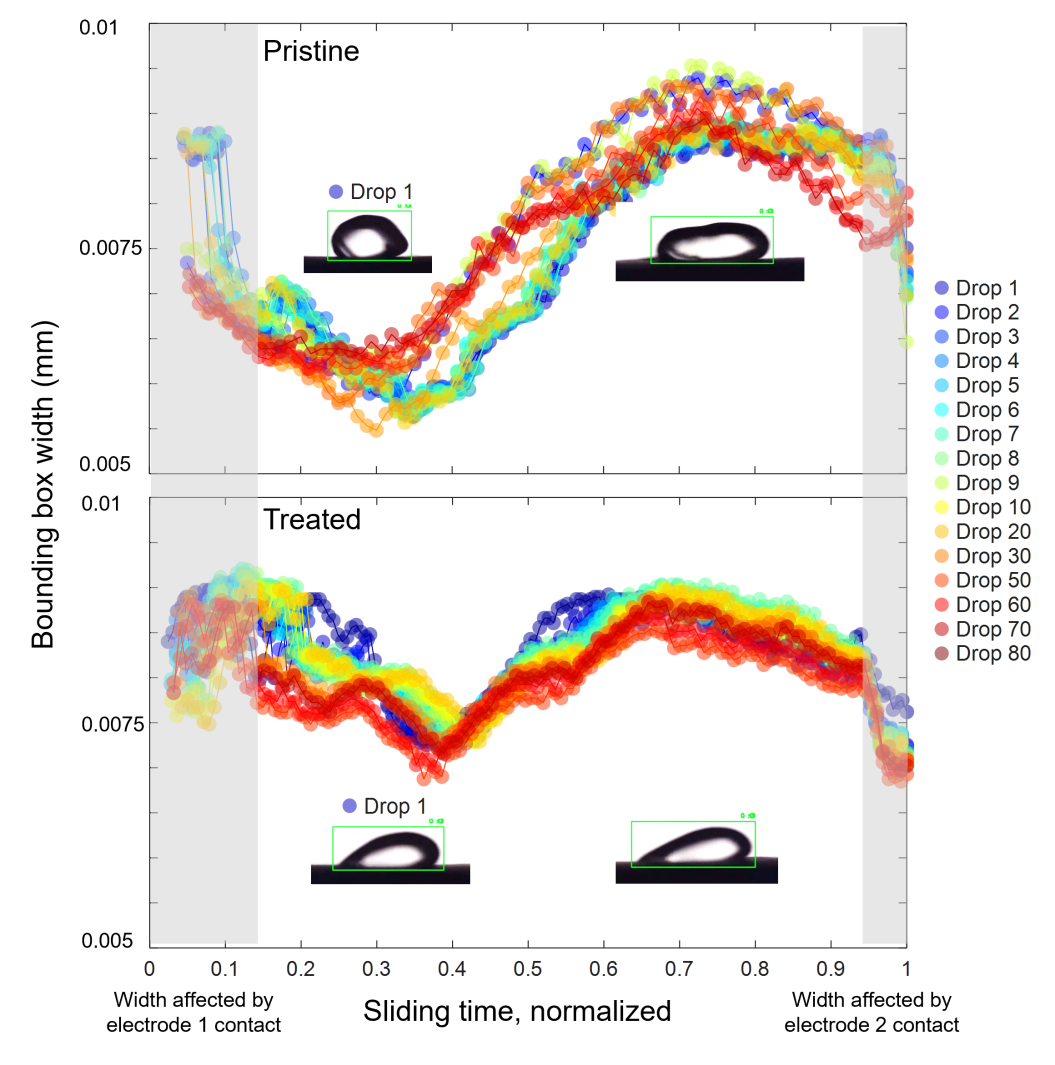}
  \caption{Droplet deformation on pristine and treated \textit{C. esculenta} leaf surface along a 40 mm slide. We used the width of the bounding box generated during the automated object tracking representing the droplet length to estimate the droplet's interaction with the surface as function of droplet number and sliding time (normalized dividing the actual time by the time needed to slide 40 mm for each droplet). The lower upper panel shows the pristine leaf and the lower panel the same sliding path after epicuticular wax modification (treated leaf). Increase in the bounding box width correlates with droplet elongation (see inset images); the gray areas indicate sliding zones in which droplet motion is partially affected by crossing the electrodes, $E_1$ and $E_2$. Excluding these zones from the analysis, a clear trend is visible for treated leaves: initial droplets, especially droplet 1, show larger changes in the width than, e.g., droplet 80, suggesting stronger attractive interactions with the surface when simultaneously also charging is highest. Video 5 and Video 6 show the droplet elongation on pristine and treated leaves, respectively.}
  \label{ref_fig7}
\end{figure}

Another feature that our automatic tracking algorithm allowed us to analyze is the observation of the width of the bounding box tracking the droplet over time. When droplets slide along a surface, their apparent base length varies, usually as expected due to three coupled effects: dynamic spreading at the advancing edge and retraction at the receding edge, pinning–depinning processes associated with contact angle hysteresis, and small shape oscillations induced by inertia. This could be measured by tracking advancing and receding contact angles at the front and back, but they also affect the whole droplet body deforming transiently during motion. Thus, the variation and extension or reduction of the bounding box width can be seen as an approximation for the droplet’s surface interaction and their local variations along the slide path. Figure \ref{ref_fig7} shows the bounding box width for droplets 1-80 sliding on a pristine leaf and the same path subsequently treated. Excluding the variations at the beginning and end of the curves highlighted in gray, regions in which droplet widths are partially affected by contact with the electrodes, a similar behavior can be seen for pristine and treated surfaces which is an initial decrease in the width followed by droplet elongation as indicated in the insets. We assume that the overall variation seen on the pristine and treated leaf is not due to charging effects but probably based on local changes on the leaf surface in either surface structure or also underlying tissue, e.g., droplets may cross veins etc. Indeed, the macroscopic curve propagation remains similar after treatment that affects the micro-nano structure but not the macroscopic leaf properties. 

However, crucial differences occur which may be attributed to charging effects and the resulting interaction of the droplet with the surface: in general, the droplets are longer on the treated surfaces as a result of the higher surface contact, as also expected by the observed decrease in the contact angle and sliding velocity. Instead on the pristine surface, the droplets tend to form a more spherical shape initially, but they elongate later probably due to the variation in leaf surface features as mentioned earlier.
However, interestingly, especially on the treated surface, a clear trend is seen as a function of droplet number, the first droplet is generally "longest" (highest box width, blue curve) and then decreases gradually with droplet number so that droplets 50-80 are "shortest" (red curves). This difference of droplet elongation can be seen in Video 5 and Video 6 for pristine and treated leaves, respectively. This again suggests an increased surface contact of first droplets that could be caused by electrostatic spreading due to the enhanced charging. In agreement, for later droplets, when effective charge accumulation decreases, also droplet width and surface contact reduce and droplets speed up pointing towards a charge-driven attractive force causing the slowdown as similarly suggested also on artificial surfaces or charged droplets in general \cite{Giglio_dropshape_2020, Li_2022}. Interestingly, on the pristine leaf, the gradual decrease in droplet width with the droplet number is not observed within the tracking resolution. This is likely due to the lower overall charging causing less disturbance in droplet motion and velocity.

\section{Discussion}

Our combination of automated object tracking from high-speed videos and in situ high-resolution current measurements on living leaves, using the superhydrophobic model organism \textit{C. esculenta}, indicates that the electrostatic forces due to spontaneous electrification significantly impact droplet motion and shedding velocity on leaves. Indeed, approximately 30 years after the mechanistic description of the Lotus effect by Barthlott and Neinhuis et al. \cite{barthlott_purity_1997}, this is the first time the crucial role of epicuticular wax structure in connection to charging and droplet motion has been demonstrated on leaves. 

Variations in the plasticity of the epicuticular wax layer leading to a smoother or less nanostructured form are common in nature and can be caused by heat-, photo-, or erosion-damage, by intrinsically different wax compositions between species, or in the same species that inhabit and respond to different environments \cite{Riederer_2006, Shepherd_StressWax_2006, schefus_climatic_2005}. We prove that charging is caused and strongly affected by the plasticity of the epicuticular wax layer, resulting in a significant 30-40 fold gain in charge generation as averaged over several identical experiments. For individual experiments the gain varied from tenfold to more than one hundredfold. When the wax layer roughness is decreased, it leads to the improvement of the effective water-surface contact for charge transfer. We observe this effect in: I) traveled distances and droplet velocity, II) charge measurements, III) samples with tuned epicuticular wax structures, and IV) droplet body deformation during sliding. All of our observations confirm that spontaneous charging due to liquid-solid contact electrification correlates with the observed dynamic variation of droplet motion on the leaf. Consequently any of the above mentioned natural variation in the wax layer will vary water shedding and residing time of droplets and affect the cascade of ecophysiological follow-up reactions.     

The similar trade-off between charge accumulation and droplet velocity have recently been reported on precisely engineered synthetic substrates (especially and almost exclusively on fluorinated surfaces \cite{dratschow_liquid_2025}), where a tunable surface topography allows for more straightforward quantification. Interestingly, the quantification of charging on leaves, as well as the estimated electrostatic forces, are very similar to those reported on synthetic substrates: the charges measured for the first droplet on pristine leaves are in the range of $Q_{\mathrm{p, D1}} = -0.02 \text{ to} -0.15 \,\text{nC}$ per $30 \,\mu\text{L}$ droplet and treated leaves, $Q_{\mathrm{t, D1}} = -2.8 \text{ to} -5.2 \,\text{nC}$. The measured charges are in the same order of magnitude or even slightly higher in the case of treated leaves, than the charge measured on droplets on artificial, fluorinated surfaces (which are typically between 1 and 4 nC, Extended Data Figure \ref{ref_figsX}) under similar experimental conditions. 

Moreover, the electrostatic force estimated from measurements on treated leaves is approximately 11~$\mu$N (30~$\mu$L droplet, 40° tilt angle, Extended Data Figure \ref{ref_figs9}), which is in a comparable regime to the electrostatic resistive forces measured on artificial fluorinated surfaces (20--80~$\mu$N, 45~$\mu$L droplet, 50° tilt angle) \cite{Li_2022} under similar experimental conditions. This not only clearly suggests that electrostatic forces affect droplet charging on leaves but also shows that the materials in the epicuticular waxes (a complex mixture of aliphatic hydrocarbons with multiple functional groups) may have a similar charge transfer capability; even though it occurs, however, likely through a different mechanism as also suggested by the polarity of droplet charge (negative on leaves, positive on fluorinated surfaces). Indeed, this raises two other interesting aspects.

In engineering, fluorinated surfaces are still the "gold standard" in droplet energy harvesting \cite{lin_contact_2022}. Our results show that simply varying the epicuticular wax structure leads to comparable droplet charging on leaves (Extended Data Figure \ref{ref_figsX}) suggesting directions to gain alternative, more sustainable bio-materials for controlling surface charging: for example, for electrostatic discharge protection (charge reduction via structures like those on pristine leaves) or for droplet energy harvesting (charge gain like that on treated leaves), for which environmentally critical fluorinated polymers are currently the state of the art. Further investigations into which structures, alongside superhydrophobicity, can reduce charge effects on droplet motion could help develop artificial materials for applications like microfluidics, where droplet charging should be minimized \cite{Mazutis2013,Li2015, Wang2025}. Further application-relevant implication of water-leaf charging is in precision agriculture  spraying where tuning charge-leaf interactions can improve droplet retention on the leaf and reduce soil contamination \cite{damak_MITspraying_2016}.   

In nature, numerous wax conformations, crystal shapes, and hierarchical assemblies exist, depending on the plant species and environmental conditions, such as humidity and temperature, as well as leaf age and potential damage. Our results clearly demonstrate that the wax layer can be tuned to significantly impact droplet motion. This raises again the question of whether some plants may have specifically evolved the wax layer to increase or decrease charging, thereby electrostatically driving leaf water transport. 

To better understand charging phenomena in nature during events such as rain or dew formation, fundamental investigations of complex materials in living plants are required. These must go beyond model surfaces to further elucidate mechanisms such as the contribution of spontaneous charging. Our combination of methodologies—providing an experimental and theoretical framework to derive parameters such as wax layer gain, resistive force calculations, and electrostatic contributions—serves as an example of successfully investigating such complex systems.    
 
\section{Conclusions}

Plants have evolved complex leaf morphologies and textures to precisely regulate how droplets wet, stick, bounce, and shed, driven also by in situ electrical phenomena next to structure and chemistry. Our results provide an experimental and theoretical toolkit for investigating droplet dynamics as function of interfacial charging on complex biological surfaces and we demonstrate that electrostatic forces play a fundamental role in the dynamics even on living leaves, both, with and without "Lotus effect"-like superhydrophobicity. Moreover, we prove that droplet motion and charge accumulation are dictated by the plasticity of the epicuticular wax layer. Specifically, we show that this layer can enhance charging by a factor of 30-40, while simultaneously reducing droplet velocity by half. In addition,
charging is fundamentally affected by the surface history and the sequence of
droplets that the surface has ”seen” before. Within 10-20 droplets, the droplets
can speed up by 25\% due to a decreased charge accumulation.

This work provides the first quantitative characterization of the intrinsic connection between charge accumulation and droplet speed on biological surfaces. Notably, the discovery that thermal treatment can render biological surfaces electrically comparable, or even superior, to synthetic fluoropolymers expands the portfolio of sustainable materials for charge tuning. These findings also raise fundamental questions regarding the evolutionary role of epicuticular waxes: specifically, whether these architectures could be optimized to exploit charges physiologically.  Ultimately, our work establishes a basis for diverse technological advancements. By bridging the gap between biological interface science and electro-hydrodynamics, these insights provide directions for enhancing droplet retention in precision agricultural spraying and developing fluorine-free, bio-inspired materials for high-efficiency droplet-based energy harvesting.

\section{Acknowledgments}

FM acknowledges funding from the European Research Council (ERC) for the project EpiC (grant agreement ID 101124721, \\ https://doi.org/10.3030/101124721). SS and MD gratefully acknowledge funding by the European Union (EU) under the Horizon Europe research and innovation programme, EIC Pathfinder - grant No. 101187428 (iNSIGHT) and from the European Research Council ERC-PoC2 grant No. 101187935 (LBFAST).

\clearpage
\bibliography{References}

@article{Wang2025,
    author = {Wang, Dongbao and Chagot, Loic and Wang, Junfeng and Angeli, Panagiota},
    title = {Effect of electric field on droplet formation in a co-flow microchannel},
    journal = {Physics of Fluids},
    volume = {37},
    number = {2},
    pages = {023331},
    year = {2025},
    month = {02},
    issn = {1070-6631},
    doi = {10.1063/5.0250687},
    url = {https://doi.org/10.1063/5.0250687},
    eprint = {https://pubs.aip.org/aip/pof/article-pdf/doi/10.1063/5.0250687/20383069/023331_1_5.0250687.pdf},
}

@Article{Mazutis2013,
author={Mazutis, Linas
and Gilbert, John
and Ung, W. Lloyd
and Weitz, David A.
and Griffiths, Andrew D.
and Heyman, John A.},
title={Single-cell analysis and sorting using droplet-based microfluidics},
journal={Nature Protocols},
year={2013},
month={May},
day={01},
volume={8},
number={5},
pages={870-891},
issn={1750-2799},
doi={10.1038/nprot.2013.046},
url={https://doi.org/10.1038/nprot.2013.046}
}

@Article{Li2015,
AUTHOR = {Li, Jingmei and Liu, Zhou and Huang, Haibo and Shum, Ho Cheung},
TITLE = {Shielding Electric Fields to Prevent Coalescence of Emulsions in Microfluidic Channels Using a 3D Metallic Coil},
JOURNAL = {Micromachines},
VOLUME = {6},
YEAR = {2015},
NUMBER = {10},
PAGES = {1459--1468},
URL = {https://www.mdpi.com/2072-666X/6/10/1430},
ISSN = {2072-666X},
DOI = {10.3390/mi6101430}
}

@misc{yolo11_ultralytics,
  author = {Jocher, Glenn and Qiu, Jing},
  title = {{Ultralytics YOLO11} (Version 11.0.0)},
  year = {2024},
  howpublished = {\url{https://github.com/ultralytics/ultralytics}},
}

@misc{droplet_data,
author = {Mihir Durve},
publisher= {IIT Dataverse},
title= {DropTrack - Droplet identification and tracking tool dataset},
year= {2024},
howpublished = {\url{https://doi.org/10.48557/PQPSN3}},
}

@article{yolo_original,
  author={J. {Redmon} and S. {Divvala} and R. {Girshick} and A. {Farhadi}},
  journal={2016 IEEE Conference on Computer Vision and Pattern Recognition (CVPR)}, 
  title={You Only Look Once: Unified, Real-Time Object Detection}, 
  year={2016},
  volume={},
  number={},
  pages={779-788},
  doi={10.1109/CVPR.2016.91}}

@article{deepsort_original,
  author={N. {Wojke} and A. {Bewley} and D. {Paulus}},
  journal={2017 IEEE International Conference on Image Processing (ICIP)}, 
  title={Simple online and realtime tracking with a deep association metric}, 
  year={2017},
  volume={},
  number={},
  pages={3645-3649},
  doi={10.1109/ICIP.2017.8296962}}

@article{durve_2022,
    author = {Durve, Mihir and Tiribocchi, Adriano and Bonaccorso, Fabio and Montessori, Andrea and Lauricella, Marco and Bogdan, Michał and Guzowski, Jan and Succi, Sauro},
    title = {DropTrack—Automatic droplet tracking with YOLOv5 and DeepSORT for microfluidic applications},
    journal = {Physics of Fluids},
    volume = {34},
    number = {8},
    pages = {082003},
    year = {2022},
    month = {08},
    issn = {1070-6631},
    doi = {10.1063/5.0097597},
    url = {https://doi.org/10.1063/5.0097597},
    eprint = {https://pubs.aip.org/aip/pof/article-pdf/doi/10.1063/5.0097597/16566378/082003_1_online.pdf},
}

@Article{Li_2022,
author={Li, Xiaomei
and Bista, Pravash
and Stetten, Amy Z.
and Bonart, Henning
and Sch{\"u}r, Maximilian T.
and Hardt, Steffen
and Bodziony, Francisco
and Marschall, Holger
and Saal, Alexander
and Deng, Xu
and Berger, R{\"u}diger
and Weber, Stefan A. L.
and Butt, Hans-J{\"u}rgen},
title={Spontaneous charging affects the motion of sliding drops},
journal={Nature Physics},
year={2022},
month={Jun},
day={01},
volume={18},
number={6},
pages={713-719},
issn={1745-2481},
doi={10.1038/s41567-022-01563-6},
url={https://doi.org/10.1038/s41567-022-01563-6}
}

@article{Sobarzo_2025,
	title = {Spontaneous ordering of identical materials into a triboelectric series},
	volume = {638},
	copyright = {2025 The Author(s)},
	issn = {1476-4687},
	url = {https://www.nature.com/articles/s41586-024-08530-6},
	doi = {10.1038/s41586-024-08530-6},
	language = {en},
	number = {8051},
	urldate = {2025-09-23},
	journal = {Nature},
	author = {Sobarzo, Juan Carlos and Pertl, Felix and Balazs, Daniel M. and Costanzo, Tommaso and Sauer, Markus and Foelske, Annette and Ostermann, Markus and Pichler, Christian M. and Wang, Yongkang and Nagata, Yuki and Bonn, Mischa and Waitukaitis, Scott},
	month = feb,
	year = {2025},
	note = {Publisher: Nature Publishing Group},
	pages = {664--669},
}

@article{armiento_liquid-solid_2022,
	title = {Liquid-solid contact electrification when water droplets hit living plant leaves},
	volume = {3},
	copyright = {2022 The Author(s)},
	issn = {2662-4443},
	url = {http://www.nature.com/articles/s43246-022-00302-x},
	doi = {10.1038/s43246-022-00302-x},
	language = {en},
	number = {1},
	urldate = {2022-12-09},
	journal = {Communications Materials},
	author = {Armiento, Serena and Filippeschi, Carlo and Meder, Fabian and Mazzolai, Barbara},
	month = oct,
	year = {2022},
	note = {Number: 1
Publisher: Nature Publishing Group},
	pages = {1--12},
}

@article{miljkovic_electrostatic_2013,
	title = {Electrostatic charging of jumping droplets},
	volume = {4},
	copyright = {2013 Springer Nature Limited},
	issn = {2041-1723},
	url = {https://www.nature.com/articles/ncomms3517},
	doi = {10.1038/ncomms3517},
	language = {en},
	number = {1},
	urldate = {2025-09-25},
	journal = {Nature Communications},
	author = {Miljkovic, Nenad and Preston, Daniel J. and Enright, Ryan and Wang, Evelyn N.},
	month = sep,
	year = {2013},
	note = {Publisher: Nature Publishing Group},
	pages = {2517},
}

@article{sun_surface_2019,
	title = {Surface charge printing for programmed droplet transport},
	volume = {18},
	copyright = {2019 The Author(s), under exclusive licence to Springer Nature Limited},
	issn = {1476-4660},
	url = {https://www.nature.com/articles/s41563-019-0440-2},
	doi = {10.1038/s41563-019-0440-2},
	language = {en},
	number = {9},
	urldate = {2025-01-28},
	journal = {Nature Materials},
	author = {Sun, Qiangqiang and Wang, Dehui and Li, Yanan and Zhang, Jiahui and Ye, Shuji and Cui, Jiaxi and Chen, Longquan and Wang, Zuankai and Butt, Hans-Jürgen and Vollmer, Doris and Deng, Xu},
	month = sep,
	year = {2019},
	note = {Publisher: Nature Publishing Group},
	pages = {936--941},
}

@inbook{Riederer_2006,
author = {Riederer, Markus},
publisher = {John Wiley \& Sons, Ltd},
isbn = {9780470988718},
title = {Introduction: Biology of the Plant Cuticle},
booktitle = {Annual Plant Reviews Volume 23: Biology of the Plant Cuticle},
chapter = {1},
pages = {1-10},
doi = {https://doi.org/10.1002/9780470988718.ch1},
url = {https://onlinelibrary.wiley.com/doi/abs/10.1002/9780470988718.ch1},
eprint = {https://onlinelibrary.wiley.com/doi/pdf/10.1002/9780470988718.ch1},
year = {2006}
}

@article{koch_multifunctional_2009,
	title = {Multifunctional surface structures of plants: {An} inspiration for biomimetics},
	volume = {54},
	issn = {0079-6425},
	shorttitle = {Multifunctional surface structures of plants},
	url = {https://www.sciencedirect.com/science/article/pii/S0079642508000704},
	doi = {10.1016/j.pmatsci.2008.07.003},
	number = {2},
	urldate = {2025-09-25},
	journal = {Progress in Materials Science},
	author = {Koch, Kerstin and Bhushan, Bharat and Barthlott, Wilhelm},
	month = feb,
	year = {2009},
	pages = {137--178},
}

@article{yeats_formation_2013,
	title = {The {Formation} and {Function} of {Plant} {Cuticles}},
	volume = {163},
	issn = {0032-0889},
	url = {https://doi.org/10.1104/pp.113.222737},
	doi = {10.1104/pp.113.222737},
	number = {1},
	journal = {Plant Physiology},
	author = {Yeats, Trevor H. and Rose, Jocelyn K.C.},
	month = jul,
	year = {2013},
	note = {\_eprint: https://academic.oup.com/plphys/article-pdf/163/1/5/36098772/plphys\_v163\_1\_5.pdf},
	pages = {5--20},
}

@article{dratschow_liquid_2025,
	title = {Liquid slide electrification: advances and open questions},
	shorttitle = {Liquid slide electrification},
	url = {https://pubs.rsc.org/en/content/articlelanding/2025/sm/d4sm01289e},
	doi = {10.1039/D4SM01289E},
	language = {en},
	urldate = {2025-02-04},
	journal = {Soft Matter},
	author = {D. Ratschow, Aaron and Butt, Hans-Jürgen and Hardt, Steffen and L. Weber, Stefan A.},
	year = {2025},
	note = {Publisher: Royal Society of Chemistry},
}

@article{yatsuzuka_electrification_1994,
	title = {Electrification phenomena of pure water droplets dripping and sliding on a polymer surface},
	volume = {32},
	issn = {0304-3886},
	url = {https://www.sciencedirect.com/science/article/pii/0304388694900051},
	doi = {10.1016/0304-3886(94)90005-1},
	number = {2},
	urldate = {2025-09-25},
	journal = {Journal of Electrostatics},
	author = {Yatsuzuka, Kyoko and Mizuno, Yukio and Asano, Kazutoshi},
	month = apr,
	year = {1994},
	pages = {157--171},
}

@article{lin_contact_2022,
	title = {Contact {Electrification} at the {Liquid}–{Solid} {Interface}},
	volume = {122},
	issn = {0009-2665},
	url = {https://doi.org/10.1021/acs.chemrev.1c00176},
	doi = {10.1021/acs.chemrev.1c00176},
	number = {5},
	urldate = {2025-01-28},
	journal = {Chemical Reviews},
	author = {Lin, Shiquan and Chen, Xiangyu and Wang, Zhong Lin},
	month = mar,
	year = {2022},
	note = {Publisher: American Chemical Society},
	pages = {5209--5232},
}

@article{burgo_where_2016,
	title = {Where is water in the triboelectric series?},
	volume = {80},
	issn = {0304-3886},
	url = {https://www.sciencedirect.com/science/article/pii/S030438861630002X},
	doi = {10.1016/j.elstat.2016.01.002},
	urldate = {2025-09-26},
	journal = {Journal of Electrostatics},
	author = {Burgo, Thiago A. L. and Galembeck, Fernando and Pollack, Gerald H.},
	month = apr,
	year = {2016},
	pages = {30--33},
}

@article{jayaprakash_enhancing_2025,
	title = {Enhancing spray retention using cloaked droplets to reduce pesticide pollution},
	volume = {21},
	issn = {1744-6848},
	url = {https://pubs.rsc.org/en/content/articlelanding/2025/sm/d4sm01496k},
	doi = {10.1039/D4SM01496K},
	language = {en},
	number = {19},
	urldate = {2025-09-26},
	journal = {Soft Matter},
	author = {Jayaprakash, Vishnu and Rufer, Simon and Panat, Sreedath and Varanasi, Kripa K.},
	month = may,
	year = {2025},
	note = {Publisher: The Royal Society of Chemistry},
	pages = {3688--3699},
}

@article{armiento_device_2023,
	title = {Device for {Simultaneous} {Wind} and {Raindrop} {Energy} {Harvesting} {Operating} on the {Surface} of {Plant} {Leaves}},
	volume = {8},
	issn = {2377-3766},
	doi = {10.1109/LRA.2023.3250006},
	number = {4},
	journal = {IEEE Robotics and Automation Letters},
	author = {Armiento, Serena and Meder, Fabian and Mazzolai, Barbara},
	month = apr,
	year = {2023},
	note = {Conference Name: IEEE Robotics and Automation Letters},
	pages = {2269--2276},
}

@article{wu_fully_2020,
	title = {Fully {Biodegradable} {Water} {Droplet} {Energy} {Harvester} {Based} on {Leaves} of {Living} {Plants}},
	volume = {12},
	issn = {1944-8244},
	url = {https://doi.org/10.1021/acsami.0c17601},
	doi = {10.1021/acsami.0c17601},
	number = {50},
	urldate = {2023-01-26},
	journal = {ACS Applied Materials \& Interfaces},
	author = {Wu, Hao and Chen, Zefeng and Xu, Guoqiang and Xu, Jianbin and Wang, Zuankai and Zi, Yunlong},
	month = dec,
	year = {2020},
	note = {Publisher: American Chemical Society},
	pages = {56060--56067},
}

@article{mukherjee_synergistic_2021,
	title = {Synergistic dispersal of plant pathogen spores by jumping-droplet condensation and wind},
	volume = {118},
	url = {https://www.pnas.org/doi/10.1073/pnas.2106938118},
	doi = {10.1073/pnas.2106938118},
	number = {34},
	urldate = {2025-09-26},
	journal = {Proceedings of the National Academy of Sciences},
	author = {Mukherjee, Ranit and Gruszewski, Hope A. and Bilyeu, Landon T. and Schmale, David G. and Boreyko, Jonathan B.},
	month = aug,
	year = {2021},
	note = {Publisher: Proceedings of the National Academy of Sciences},
	pages = {e2106938118},
}

@article{kim_vortex-induced_2019,
	title = {Vortex-induced dispersal of a plant pathogen by raindrop impact},
	volume = {116},
	url = {https://www.pnas.org/doi/10.1073/pnas.1820318116},
	doi = {10.1073/pnas.1820318116},
	number = {11},
	urldate = {2025-09-26},
	journal = {Proceedings of the National Academy of Sciences},
	author = {Kim, Seungho and Park, Hyunggon and Gruszewski, Hope A. and Schmale, David G. and Jung, Sunghwan},
	month = mar,
	year = {2019},
	note = {Publisher: Proceedings of the National Academy of Sciences},
	pages = {4917--4922},
}

@article{li_sparking_2023,
	title = {Sparking potential over 1200 {V} by a falling water droplet},
	volume = {9},
	url = {https://www.science.org/doi/10.1126/sciadv.adi2993},
	doi = {10.1126/sciadv.adi2993},
	number = {46},
	urldate = {2025-07-08},
	journal = {Science Advances},
	author = {Li, Luxian and Li, Xuemei and Deng, Wei and Shen, Chun and Chen, Xinhai and Sheng, Han and Wang, Xiang and Zhou, Jianxin and Li, Jidong and Zhu, Yinlong and Zhang, Zhuhua and Yin, Jun and Guo, Wanlin},
	month = nov,
	year = {2023},
	note = {Publisher: American Association for the Advancement of Science},
	pages = {eadi2993},
}

@article{bista_high_2023,
	title = {High {Voltages} in {Sliding} {Water} {Drops}},
	volume = {14},
	url = {https://doi.org/10.1021/acs.jpclett.3c02864},
	doi = {10.1021/acs.jpclett.3c02864},
	number = {49},
	urldate = {2025-02-17},
	journal = {The Journal of Physical Chemistry Letters},
	author = {Bista, Pravash and Ratschow, Aaron D. and Butt, Hans-Jürgen and Weber, Stefan A. L.},
	month = dec,
	year = {2023},
	note = {Publisher: American Chemical Society},
	pages = {11110--11116},
}

@article{park_dynamics_2020,
	title = {Dynamics of splashed droplets impacting wheat leaves treated with a fungicide},
	volume = {17},
	url = {https://royalsocietypublishing.org/doi/10.1098/rsif.2020.0337},
	doi = {10.1098/rsif.2020.0337},
	number = {168},
	urldate = {2025-09-26},
	journal = {Journal of The Royal Society Interface},
	author = {Park, Hyunggon and Kim, Seungho and Gruszewski, Hope A. and Schmale, David G. and Boreyko, Jonathan B. and Jung, Sunghwan},
	month = jul,
	year = {2020},
	note = {Publisher: Royal Society},
	pages = {20200337},
}

@article{lacks_long-standing_2019,
	title = {Long-standing and unresolved issues in triboelectric charging},
	volume = {3},
	copyright = {2019 Springer Nature Limited},
	issn = {2397-3358},
	url = {http://www.nature.com/articles/s41570-019-0115-1},
	doi = {10.1038/s41570-019-0115-1},
	language = {en},
	number = {8},
	urldate = {2022-12-09},
	journal = {Nature Reviews Chemistry},
	author = {Lacks, Daniel J. and Shinbrot, Troy},
	month = aug,
	year = {2019},
	note = {Number: 8
Publisher: Nature Publishing Group},
	pages = {465--476},
}

@article{mertcan_ozel_why_2020,
	title = {Why {Does} {Wood} {Not} {Get} {Contact} {Charged}? {Lignin} as an {Antistatic} {Additive} for {Common} {Polymers}},
	volume = {32},
	url = {https://pubs-acs-org.biblio.iit.it/doi/10.1021/acs.chemmater.0c02421},
	doi = {https://doi.org/10.1021/acs.chemmater.0c02421},
	number = {17},
	urldate = {2022-12-12},
	journal = {Chemistry of Materials},
	author = {{Mertcan Özel} and {Fatma Demir} and {Aizimaiti Aikebaier} and {Joanna Kwiczak-Yiğitbaşı} and {H. Tarik Baytekin} and {Bilge Baytekin}},
	month = aug,
	year = {2020},
	pages = {7438--7444},
}

@article{choi_spontaneous_2017,
	title = {Spontaneous occurrence of liquid-solid contact electrification in nature: {Toward} a robust triboelectric nanogenerator inspired by the natural lotus leaf},
	volume = {36},
	issn = {22112855},
	shorttitle = {Spontaneous occurrence of liquid-solid contact electrification in nature},
	url = {https://linkinghub.elsevier.com/retrieve/pii/S2211285517302306},
	doi = {10.1016/j.nanoen.2017.04.026},
	language = {en},
	urldate = {2023-01-26},
	journal = {Nano Energy},
	author = {Choi, Dongwhi and Kim, Do Wan and Yoo, Donghyeon and Cha, Kyoung Je and La, Moonwoo and Kim, Dong Sung},
	month = jun,
	year = {2017},
	pages = {250--259},
}

@article{lenz_ecological_2022,
	title = {An ecological perspective on water shedding from leaves},
	volume = {73},
	issn = {0022-0957},
	url = {https://doi.org/10.1093/jxb/erab479},
	doi = {10.1093/jxb/erab479},
	number = {4},
	urldate = {2025-09-29},
	journal = {Journal of Experimental Botany},
	author = {Lenz, Anne-Kristin and Bauer, Ulrike and Ruxton, Graeme D},
	month = feb,
	year = {2022},
	pages = {1176--1189},
}

@article{neinhuis_characterization_1997,
	title = {Characterization and {Distribution} of {Water}-repellent, {Self}-cleaning {Plant} {Surfaces}},
	volume = {79},
	issn = {0305-7364},
	url = {https://www.sciencedirect.com/science/article/pii/S0305736497904005},
	doi = {10.1006/anbo.1997.0400},
	number = {6},
	urldate = {2025-09-29},
	journal = {Annals of Botany},
	author = {Neinhuis, C. and Barthlott, W.},
	month = jun,
	year = {1997},
	pages = {667--677},
}

@article{Kamare_wax_2025,
author = {Kamare, Behnam and Shahabi, Mahla and de Resmini, Matteo Carpi and Fernandes, Tiago and Meder, Fabian},
title = {Bio-Based Wax Interfaces for Droplet Energy Harvesting at Fluoropolymer-Like Output Levels},
journal = {Advanced Science},
volume = {n/a},
number = {n/a},
pages = {e15266},
year = {2025},
doi = {https://doi.org/10.1002/advs.202515266},
url = {https://advanced.onlinelibrary.wiley.com/doi/abs/10.1002/advs.202515266},
eprint = {https://advanced.onlinelibrary.wiley.com/doi/pdf/10.1002/advs.202515266}
}

@article{barthlott_purity_1997,
	title = {Purity of the sacred lotus, or escape from contamination in biological surfaces},
	volume = {202},
	issn = {1432-2048},
	url = {https://doi.org/10.1007/s004250050096},
	doi = {10.1007/s004250050096},
	language = {en},
	number = {1},
	urldate = {2025-12-03},
	journal = {Planta},
	author = {Barthlott, W. and Neinhuis, C.},
	month = apr,
	year = {1997},
	pages = {1--8},
}

@Article{Nishimoto_BioinspiredSelfCleaning_2012,
author ="Nishimoto, Shunsuke and Bhushan, Bharat",
title  ="Bioinspired self-cleaning surfaces with superhydrophobicity{,} superoleophobicity{,} and superhydrophilicity",
journal  ="RSC Adv.",
year  ="2013",
volume  ="3",
issue  ="3",
pages  ="671-690",
publisher  ="The Royal Society of Chemistry",
doi  ="10.1039/C2RA21260A",
url  ="http://dx.doi.org/10.1039/C2RA21260A"}

@article{Guittard_BioinspWetting_2025,
author = {Guittard, Fr{\'e}d{\'e}ric and Amigoni, Sonia and Darmanin, Thierry},
title = {Bioinspired and Biomimetic Wetting Properties},
journal = {ACS Nano},
volume = {19},
number = {41},
pages = {36005-36026},
year = {2025},
doi = {10.1021/acsnano.5c04840},
    note ={PMID: 41064970},
URL = {https://doi.org/10.1021/acsnano.5c04840},
eprint = {https://doi.org/10.1021/acsnano.5c04840}
}

@article{Dunkerley_Static_2023,
author = {Dunkerley, David},
title = {Leaf water shedding: Moving away from assessments based on static contact angles, and a new device for observing dynamic droplet roll-off behaviour},
journal = {Methods in Ecology and Evolution},
volume = {14},
number = {12},
pages = {3047-3054},
doi = {https://doi.org/10.1111/2041-210X.14253},
url = {https://besjournals.onlinelibrary.wiley.com/doi/abs/10.1111/2041-210X.14253},
eprint = {https://besjournals.onlinelibrary.wiley.com/doi/pdf/10.1111/2041-210X.14253},
year = {2023}
}

@article{Snoeijer_MovingContactLines_2013,
   author = "Snoeijer, Jacco H. and Andreotti, Bruno",
   title = "Moving Contact Lines: Scales, Regimes, and Dynamical Transitions", 
   journal= "Annual Review of Fluid Mechanics",
   year = "2013",
   volume = "45",
   number = "Volume 45, 2013",
   pages = "269-292",
   doi = "https://doi.org/10.1146/annurev-fluid-011212-140734",
   url = "https://www.annualreviews.org/content/journals/10.1146/annurev-fluid-011212-140734",
   publisher = "Annual Reviews",
   issn = "1545-4479",
   type = "Journal Article",
  }

@article{Tadmore_OpenProblems_2021,
author = {Tadmor, Rafael},
title = {Open Problems in Wetting Phenomena: Pinning Retention Forces},
journal = {Langmuir},
volume = {37},
number = {21},
pages = {6357-6372},
year = {2021},
doi = {10.1021/acs.langmuir.0c02768},
    note ={PMID: 34008988},

URL = { 
    
        https://doi.org/10.1021/acs.langmuir.0c02768
    
    

},
eprint = { 
    
        https://doi.org/10.1021/acs.langmuir.0c02768
    
    

}
}

@article{Thomson_ElectricityDrops_1894,
author = {J. J. Thomson},
title = {XXXI. On the electricity of drops },
journal = {The London, Edinburgh, and Dublin Philosophical Magazine and Journal of Science},
volume = {37},
number = {227},
pages = {341--358},
year = {1894},
publisher = {Taylor \& Francis},
doi = {10.1080/14786449408620555},


URL = { 
    
        https://doi.org/10.1080/14786449408620555
    
    

},
eprint = { 
    
        https://doi.org/10.1080/14786449408620555
    
    

}

}

@article{Xu_TriboelectricWetting_2022,
author = {Wanghuai Xu  and Yuankai Jin  and Wanbo Li  and Yuxin Song  and Shouwei Gao  and Baoping Zhang  and Lili Wang  and Miaomiao Cui  and Xiantong Yan  and Zuankai Wang },
title = {Triboelectric wetting for continuous droplet transport},
journal = {Science Advances},
volume = {8},
number = {51},
pages = {eade2085},
year = {2022},
doi = {10.1126/sciadv.ade2085},
URL = {https://www.science.org/doi/abs/10.1126/sciadv.ade2085},
eprint = {https://www.science.org/doi/pdf/10.1126/sciadv.ade2085}}

@article{wong_pitcher_2011,
	title = {Bioinspired self-repairing slippery surfaces with pressure-stable omniphobicity},
	volume = {477},
	copyright = {2011 Springer Nature Limited},
	issn = {1476-4687},
	url = {https://www.nature.com/articles/nature10447},
	doi = {10.1038/nature10447},
	language = {en},
	number = {7365},
	urldate = {2026-01-15},
	journal = {Nature},
	author = {Wong, Tak-Sing and Kang, Sung Hoon and Tang, Sindy K. Y. and Smythe, Elizabeth J. and Hatton, Benjamin D. and Grinthal, Alison and Aizenberg, Joanna},
	month = sep,
	year = {2011},
	note = {Publisher: Nature Publishing Group},
	pages = {443--447},
}

@article{kim_salvinia_2022,
	title = {Fabrication of {Salvinia}-inspired surfaces for hydrodynamic drag reduction by capillary-force-induced clustering},
	volume = {13},
	copyright = {2022 The Author(s)},
	issn = {2041-1723},
	url = {https://www.nature.com/articles/s41467-022-32919-4},
	doi = {10.1038/s41467-022-32919-4},
	language = {en},
	number = {1},
	urldate = {2026-01-15},
	journal = {Nature Communications},
	author = {Kim, Minsu and Yoo, Seunghoon and Jeong, Hoon Eui and Kwak, Moon Kyu},
	month = sep,
	year = {2022},
	note = {Publisher: Nature Publishing Group},
	pages = {5181},
}

@article{Tiribocchi_Boltzmann_2025,
title = {Lattice Boltzmann simulations for soft flowing matter},
journal = {Physics Reports},
volume = {1105},
pages = {1-52},
year = {2025},
note = {Lattice Boltzmann simulations for soft flowing matter},
issn = {0370-1573},
doi = {https://doi.org/10.1016/j.physrep.2024.11.002},
url = {https://www.sciencedirect.com/science/article/pii/S0370157324003831},
author = {Adriano Tiribocchi and Mihir Durve and Marco Lauricella and Andrea Montessori and Jean-Michel Tucny and Sauro Succi}
}

@article{Falcucci_extreme_2021,
	title = {Extreme flow simulations reveal skeletal adaptations of deep-sea sponges},
	volume = {595},
	copyright = {2021 The Author(s), under exclusive licence to Springer Nature Limited},
	issn = {1476-4687},
	url = {https://www.nature.com/articles/s41586-021-03658-1},
	doi = {10.1038/s41586-021-03658-1},
	language = {en},
	number = {7868},
	urldate = {2026-01-19},
	journal = {Nature},
	author = {Falcucci, Giacomo and Amati, Giorgio and Fanelli, Pierluigi and Krastev, Vesselin K. and Polverino, Giovanni and Porfiri, Maurizio and Succi, Sauro},
	month = jul,
	year = {2021},
	note = {Publisher: Nature Publishing Group},
	pages = {537--541},
}

@article{Giglio_dropshape_2020,
  title = {Influence of the viscosity and charge mobility on the shape deformation of critically charged droplets},
  author = {Giglio, E. and Rangama, J. and Guillous, S. and Le Cornu, T.},
  journal = {Phys. Rev. E},
  volume = {101},
  issue = {1},
  pages = {013105},
  numpages = {11},
  year = {2020},
  month = {Jan},
  publisher = {American Physical Society},
  doi = {10.1103/PhysRevE.101.013105},
  url = {https://link.aps.org/doi/10.1103/PhysRevE.101.013105}
}

@article{Durve_2024_POF,
    author = {Durve, Mihir and Orsini, Sibilla and Tiribocchi, Adriano and Montessori, Andrea and Tucny, Jean-Michel and Lauricella, Marco and Camposeo, Andrea and Pisignano, Dario and Succi, Sauro},
    title = {Measuring arrangement and size distributions of flowing droplets in microchannels through deep learning using DropTrack},
    journal = {Physics of Fluids},
    volume = {36},
    number = {2},
    pages = {022105},
    year = {2024},
    month = {02},
    issn = {1070-6631},
    doi = {10.1063/5.0185350},
    url = {https://doi.org/10.1063/5.0185350},
    eprint = {https://pubs.aip.org/aip/pof/article-pdf/doi/10.1063/5.0185350/20304298/022105_1_5.0185350.pdf},
}

@Article{Durve2023,
author={Durve, Mihir
and Orsini, Sibilla
and Tiribocchi, Adriano
and Montessori, Andrea
and Tucny, Jean-Michel
and Lauricella, Marco
and Camposeo, Andrea
and Pisignano, Dario
and Succi, Sauro},
title={Benchmarking YOLOv5 and YOLOv7 models with DeepSORT for droplet tracking applications},
journal={The European Physical Journal E},
year={2023},
month={May},
day={08},
volume={46},
number={5},
pages={32},
issn={1292-895X},
doi={10.1140/epje/s10189-023-00290-x},
url={https://doi.org/10.1140/epje/s10189-023-00290-x}
}

@article{damak_MITspraying_2016,
	title = {Enhancing droplet deposition through in-situ precipitation},
	volume = {7},
	copyright = {2016 The Author(s)},
	issn = {2041-1723},
	url = {https://www.nature.com/articles/ncomms12560},
	doi = {10.1038/ncomms12560},
	language = {en},
	number = {1},
	urldate = {2026-01-25},
	journal = {Nature Communications},
	publisher = {Nature Publishing Group},
	author = {Damak, Maher and Hyder, Md Nasim and Varanasi, Kripa K.},
	month = aug,
	year = {2016},
	pages = {12560},
}

@article{zhou_deposition_2025,
	title = {Spontaneous {Charging} from {Sliding} {Water} {Drops} {Determines} the {Interfacial} {Deposition} of {Charged} {Solutes}},
	volume = {37},
	copyright = {© 2025 The Author(s). Advanced Materials published by Wiley-VCH GmbH},
	issn = {1521-4095},
	url = {https://onlinelibrary.wiley.com/doi/abs/10.1002/adma.202420263},
	doi = {10.1002/adma.202420263},
	language = {en},
	number = {16},
	urldate = {2026-01-25},
	journal = {Advanced Materials},
	author = {Zhou, Xiaoteng and Ji, Yuwen and Ni, Zhongyuan and Lopez, Javier Garcia and Peneva, Kalina and Jiang, Shan and Knorr, Nikolaus and Berger, Rüdiger and Koynov, Kaloian and Butt, Hans-Jürgen},
	year = {2025},
	note = {\_eprint: https://advanced.onlinelibrary.wiley.com/doi/pdf/10.1002/adma.202420263},
	pages = {2420263},
}

@article{stetten_slide_2019,
	title = {Slide electrification: charging of surfaces by moving water drops},
	volume = {15},
	issn = {1744-6848},
	shorttitle = {Slide electrification},
	url = {https://pubs.rsc.org/en/content/articlelanding/2019/sm/c9sm01348b},
	doi = {10.1039/C9SM01348B},
	language = {en},
	number = {43},
	urldate = {2026-01-25},
	journal = {Soft Matter},
	publisher = {The Royal Society of Chemistry},
	author = {Stetten, Amy Z. and Golovko, Dmytro S. and Weber, Stefan A. L. and Butt, Hans-Jürgen},
	month = nov,
	year = {2019},
	pages = {8667--8679},
}

@article{Chen_irriversible_2025,
  title = {Irreversible Charging Caused by Energy Dissipation from Depinning of Droplets on Polymer Surfaces},
  author = {Chen, Shuaijia and Leon, Ronald T. and Qambari, Rahmat and Yan, Yan and Chen, Menghan and Sherrell, Peter C. and Ellis, Amanda V. and Berry, Joseph D.},
  journal = {Phys. Rev. Lett.},
  volume = {134},
  issue = {10},
  pages = {104002},
  numpages = {6},
  year = {2025},
  month = {Mar},
  publisher = {American Physical Society},
  doi = {10.1103/PhysRevLett.134.104002},
  url = {https://link.aps.org/doi/10.1103/PhysRevLett.134.104002}
}

@article{hinduja_slide_2024,
	title = {Slide electrification of drops at low velocities},
	volume = {20},
	issn = {1744-6848},
	url = {https://pubs.rsc.org/en/content/articlelanding/2024/sm/d4sm00019f},
	doi = {10.1039/D4SM00019F},
	language = {en},
	number = {15},
	urldate = {2026-01-25},
	journal = {Soft Matter},
	publisher = {The Royal Society of Chemistry},
	author = {Hinduja, Chirag and Butt, Hans-Jürgen and Berger, Rüdiger},
	month = apr,
	year = {2024},
	pages = {3349--3358},
}

@article{Eglinton_Cuticle_1967,
author = {Geoffrey Eglinton  and Richard J. Hamilton },
title = {Leaf Epicuticular Waxes},
journal = {Science},
volume = {156},
number = {3780},
pages = {1322-1335},
year = {1967},
doi = {10.1126/science.156.3780.1322},
URL = {https://www.science.org/doi/abs/10.1126/science.156.3780.1322},
eprint = {https://www.science.org/doi/pdf/10.1126/science.156.3780.1322}}

@article{Shepherd_StressWax_2006,
author = {Shepherd, Tom and Wynne Griffiths, D.},
title = {The effects of stress on plant cuticular waxes},
journal = {New Phytologist},
volume = {171},
number = {3},
pages = {469-499},
doi = {https://doi.org/10.1111/j.1469-8137.2006.01826.x},
url = {https://nph.onlinelibrary.wiley.com/doi/abs/10.1111/j.1469-8137.2006.01826.x},
eprint = {https://nph.onlinelibrary.wiley.com/doi/pdf/10.1111/j.1469-8137.2006.01826.x},
year = {2006}
}

@article{schefus_climatic_2005,
    title   = {Climatic controls on central {African} hydrology during the past 20,000 years},
    volume  = {437},
    issn    = {1476-4687},
    url     = {https://doi.org/10.1038/nature03945},
    doi     = {10.1038/nature03945},
    number  = {7061},
    journal = {Nature},
    author  = {Schefu{\ss}, Enno and Schouten, Stefan and Schneider, Ralph R.},
    year    = {2005},
    pages   = {1003--1006}
}

@article{kreder_anti-icing_2016,
	title = {Design of anti-icing surfaces: smooth, textured or slippery?},
	volume = {1},
	copyright = {2016 Macmillan Publishers Limited},
	issn = {2058-8437},
	shorttitle = {Design of anti-icing surfaces},
	url = {https://www.nature.com/articles/natrevmats20153},
	doi = {10.1038/natrevmats.2015.3},
	language = {en},
	number = {1},
	urldate = {2026-01-26},
	journal = {Nature Reviews Materials},
	publisher = {Nature Publishing Group},
	author = {Kreder, Michael J. and Alvarenga, Jack and Kim, Philseok and Aizenberg, Joanna},
	month = jan,
	year = {2016},
	pages = {15003},
}
\bibliographystyle{ieeetr}

\clearpage 

\newpage
\setcounter{figure}{0} 
\renewcommand{\thefigure}{S\arabic{figure}} \renewcommand{\theHequation}{\thesection.\arabic{equation}} 
\renewcommand{\theHfigure}{SM.\arabic{figure}}

\renewcommand{\figurename}{Extended Data Figure}
\setcounter{figure}{0} 
\renewcommand{\thefigure}{\arabic{figure}} 

\pagebreak

\subsection{Extended figure: Droplet motion of different leaves}

\begin{figure}[htbp!]
  \centering
  \includegraphics[width=\textwidth]{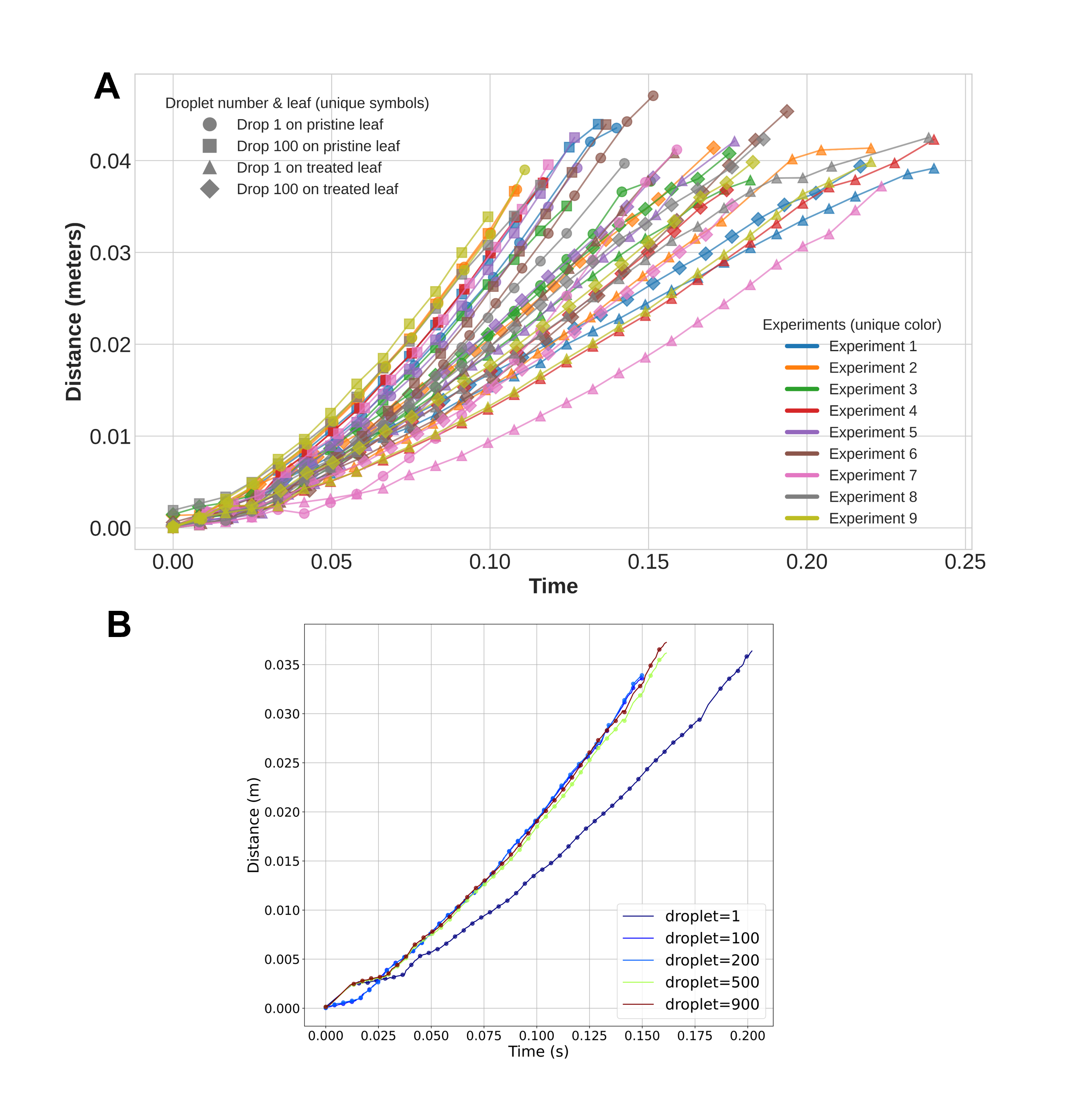}
  \caption{Additional data of droplet dynamics measured across independent experiments on 9 different leaves. a) All measurements were conducted across four independent experiments (Experiment 1–9), with each one represented by a unique color to highlight experimental reproducibility under identical ambient conditions. The time axis starts at t=0 at the moment of first detection for each droplet. Data points represent every tenth measured value for clarity. b) Results for spatiotemporal droplet dynamics on treated leaves and higher droplet numbers comparing droplet D1 with D100, D200, D500, D900 showing that beyond D100 no significant furhter variation in the dynamics have been observed also for treated leaves.}
  \label{ref_figs2}
\end{figure}

\pagebreak
\subsection{Extended figure: Surface roughness amplitudes on pristine and treated leaves}
\begin{figure}[htbp!]
  \centering
  \includegraphics[width=\textwidth]{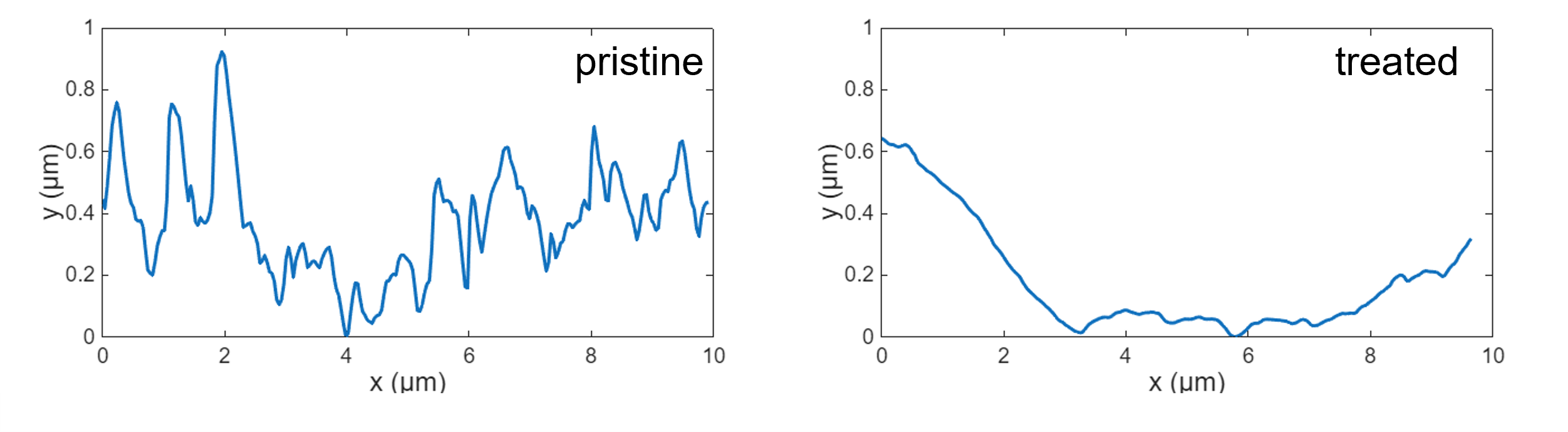}
  \caption{Surface roughness amplitude on pristine and treated leaves. AFM surface topography profiles have been recorded at the same position on a living \textit{Colocasia esculenta} leaf, before (pristine, left panel) and after modification of the epicuticular wax layer (treated, right panel). This confirms surface smoothing of the micro-nano structure of the wax crystals through the gentle thermal treatment melting the wax crystals into a smoother layer as also suggested by the increased wettability of the treated surface.}
  \label{ref_figs3}
\end{figure}

\pagebreak
\subsection{Extended figure: Comparison between droplet motion and a material body sliding on frictionless inclined planes }

\begin{figure}[h!]
  \centering
  \includegraphics[width=0.75\textwidth]{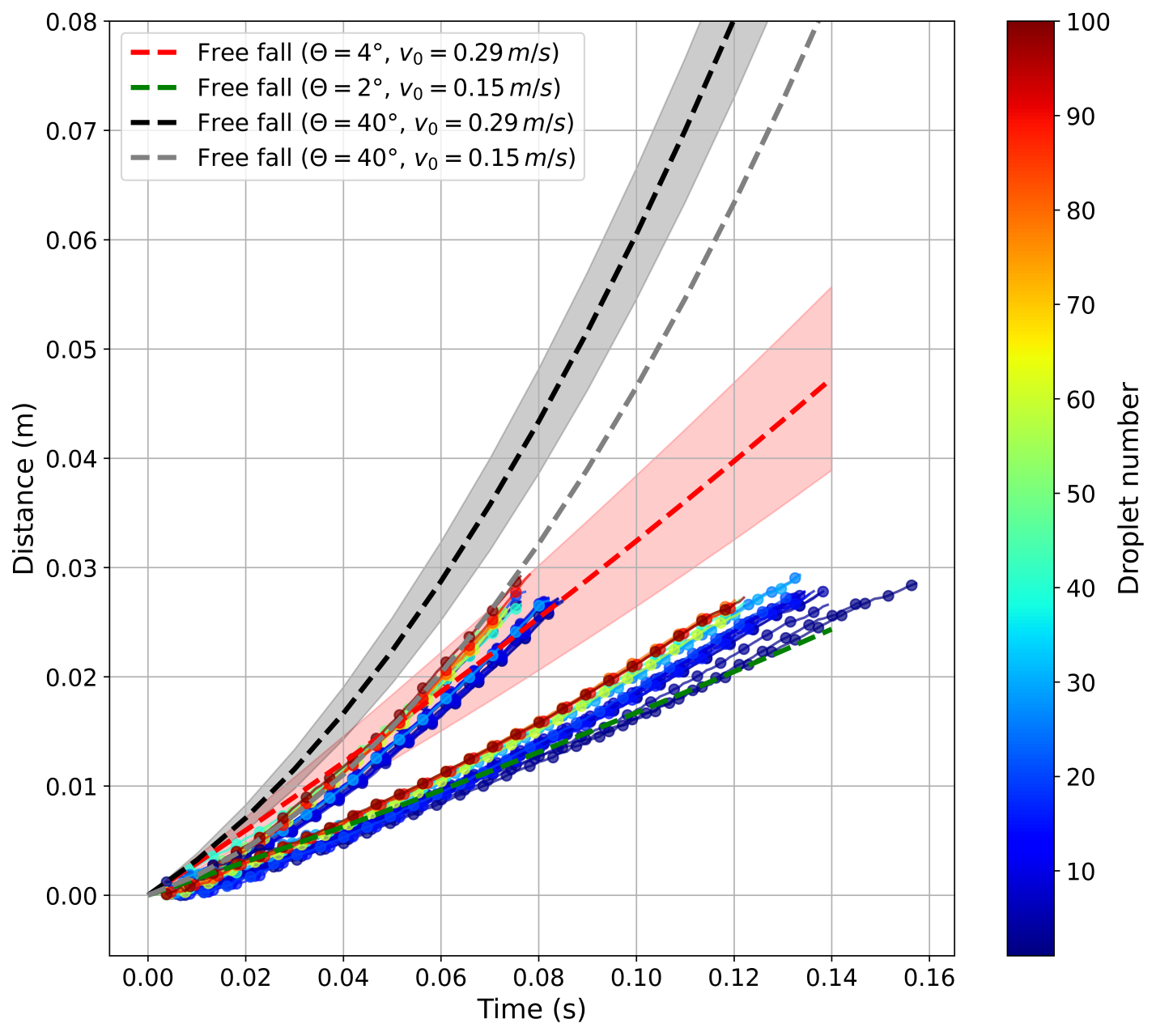}
  \caption{\textbf{Comparison of droplet dynamics on pristine and surface-treated leaves, alongside the trajectory of an object sliding on a frictionless and inclined surface under gravity.} Droplet trajectories (solid lines with points, color-coded by droplet number) are benchmarked against idealized frictionless sliding surfaces with different tilt-angle (dashed lines). The black dashed line represents a theoretical frictionless trajectory at the experimental tilt of $40^{\circ}$ ($v_0 = 0.29$~m/s). The bands illustrate the variance in trajectory evolution, centered at $0.29$ m/s with a deviation of $\pm 0.06$ m/s.  Interfacial resistance effectively renormalizes the effective tilt-angle: pristine leaf trajectories (upper cluster) align with a $4^{\circ}$ frictionless incline (red dashed line), while treated surface trajectories (lower cluster) match a $2^{\circ}$ incline with a reduced initial velocity ($v_0 = 0.15$~m/s, green dashed line). This velocity reduction from $0.29$~m/s to $0.15$~m/s extends liquid-solid contact time, yielding an approximately 30-40 fold gain (We observed even upto 100-fold gain on rare occasions) in charge accumulation. Details of the resistive force are provided in Supporting Information Section \ref{calc_resistive_forces}. \label{ref_figs4}}
\end{figure}

\pagebreak

\subsection{Extended figure: Slide-induced droplet charging on individual replicates of pristine and treated leaves}
\begin{figure}[htbp!]
  \centering
  \includegraphics[width=\textwidth]{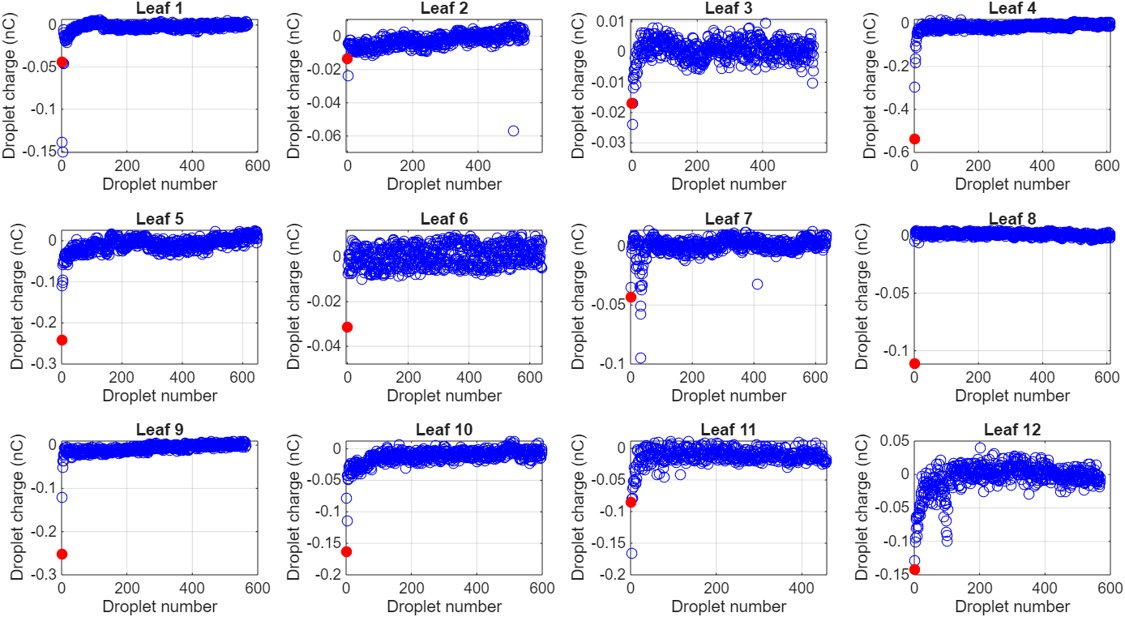}
  \caption{Pristine leaves: Analysis of charges accumulated in the droplets during a 40 mm slide as function of droplet number for 12 different biological replicates (n = 12 leaves from different plants). Each panel shows results from an individual pristine leaf and each data point represents a single droplet, the first droplet is highlighted with a red marker.}
  \label{ref_figs5}
\end{figure}
\FloatBarrier

\begin{figure}[htbp!]
  \centering
  \includegraphics[width=\textwidth]{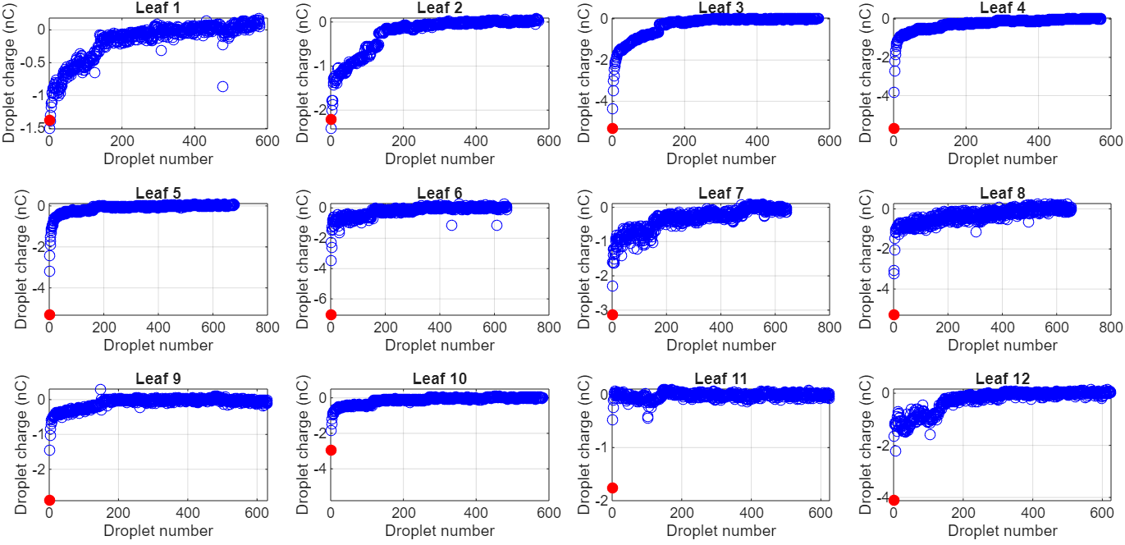}
  \caption{Treated leaves: Analysis of charges accumulated in the droplets during a 40 mm slide as function of droplet number for 12 different biological replicates (n = 12 leaves from different plants). Each panel shows results from an individual treated leaf and each data point represents a single droplet, the first droplet is highlighted with a red marker. The leaf number indicates which is the corresponding pristine surface for which charge measurements are given in Extended Data Figure \ref{ref_figs5}.}
  \label{ref_figs6}
\end{figure}
\FloatBarrier

\pagebreak
\subsection{Extended figure: Gain in surface charging caused by the treatment}
\begin{figure}[htbp!]
  \centering
  \includegraphics[width=0.75\textwidth]{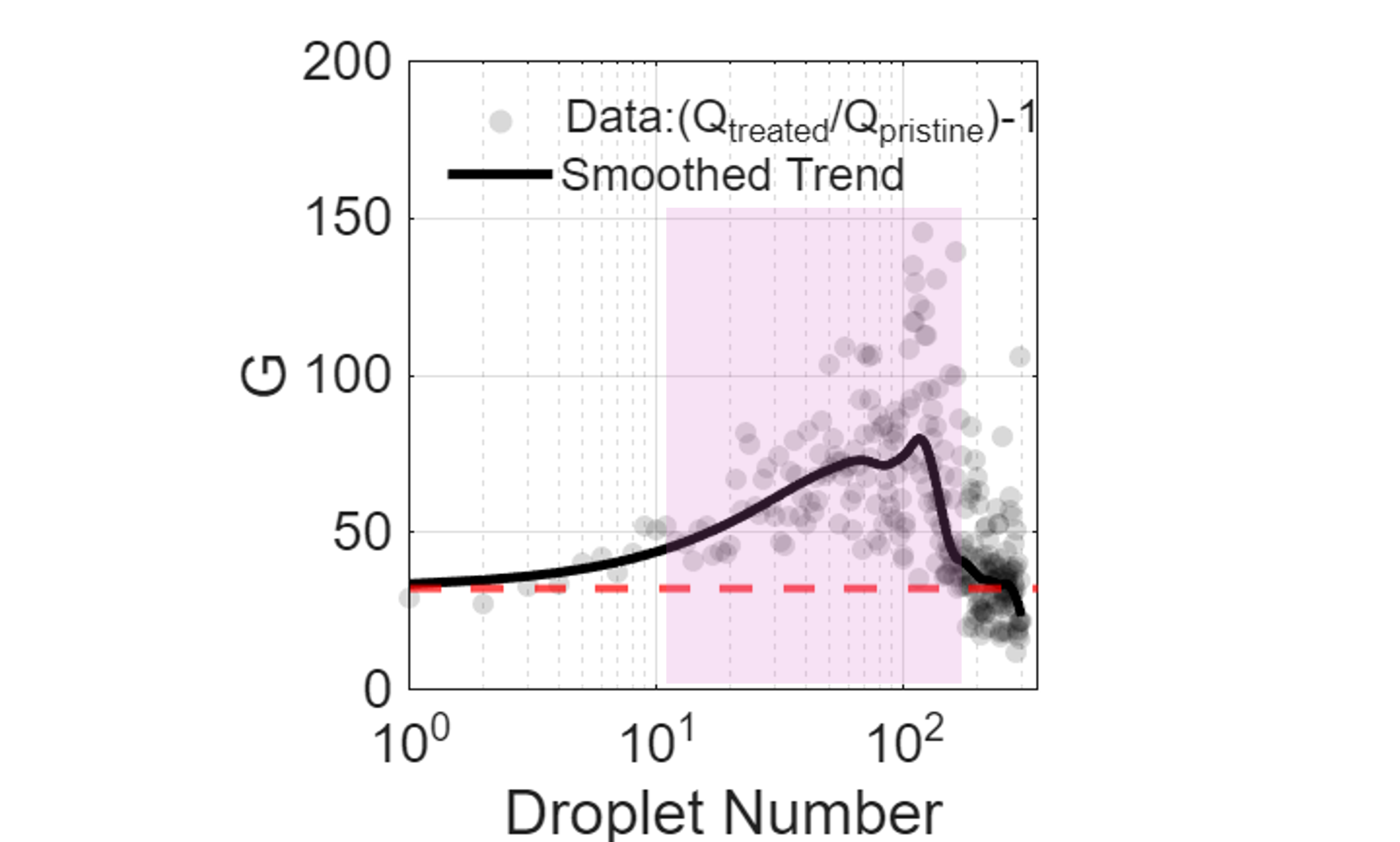} 
  \caption{The modification of the epicuticular wax layer leads to a significant gain in the surface charge that has been calculated by $G = (Q_{treated} / Q_{pristine})-1$ where $Q_{treated}$ and $Q_{pristine}$ are the droplet charges as function of droplet number on treated and pristine leaves, respectively. An about $\approx 35$-fold gain after treatment was observed, both, for initial droplets D1 to D10 sliding on the surfaces where typically highest charging was observed (flat start of smoothed data curve shown as black line) as well as for droplets $>$100-200 where charge formation reached saturation (indicated by dashed red line). The higher gain observed in the transition zone (highlighted in pink) may be attributed to the asynchronous saturation rates, where the charges saturate faster on the pristine leaves than on the treated surface. The overall 30-40 fold gain is likely due to a more effective liquid-solid contact, resulting from the reduction of the superhydrophobic wax conformation to a smoother plasticity with more charge-transfer sites.}
  \label{ref_figs7}
\end{figure}

\pagebreak
\subsection{Extended figure: Comparison of droplet charging on leaves and artificial surfaces}
\begin{figure}[htbp!]
  \centering
  \includegraphics[width=1\textwidth]{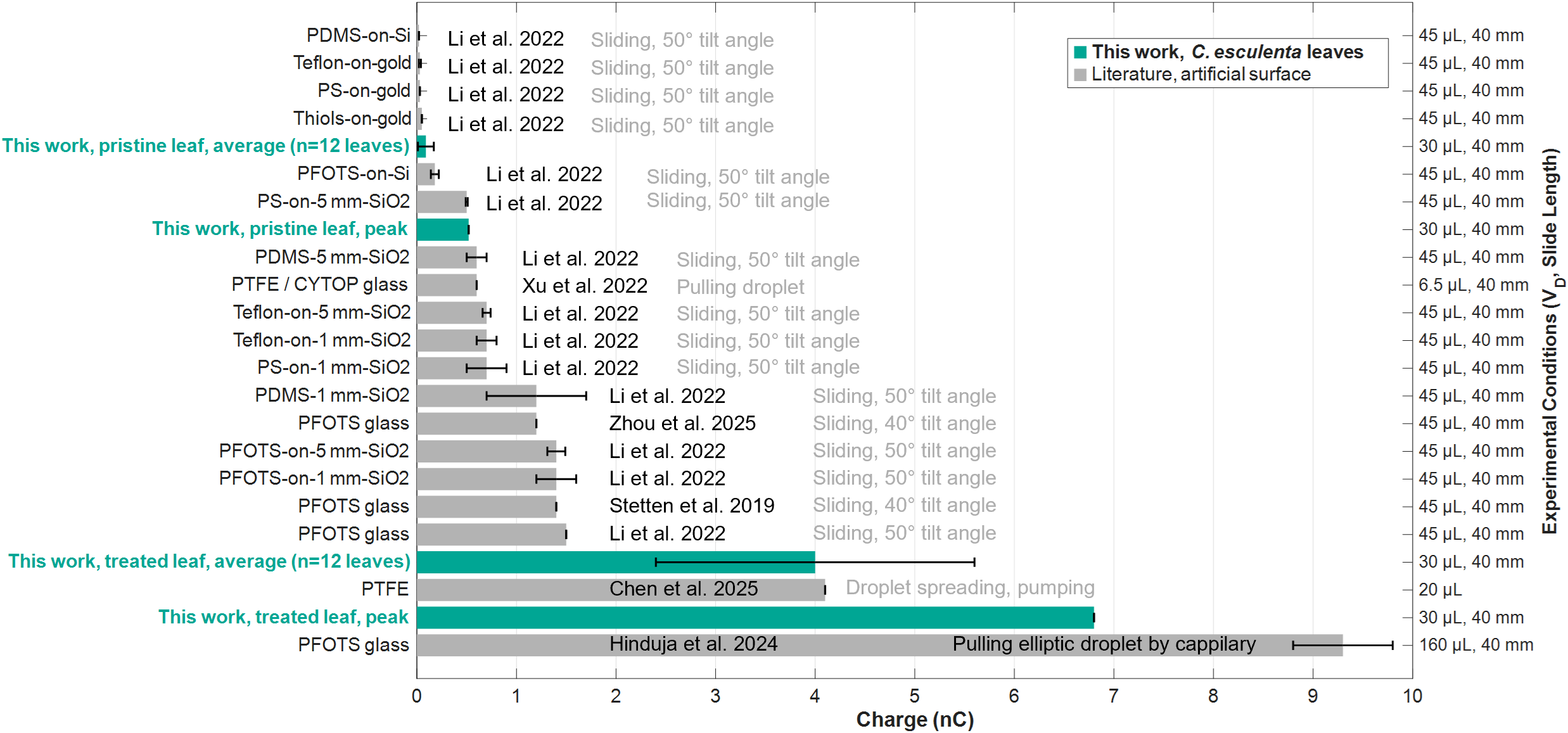} 
  \caption{Comparison of droplet charging on leaves and artificial surfaces (mainly perfluorinated, silanized glass or gold surfaces). The literature references have been chosen due to similarity of the analytical approach that was used to obtain the data to render the comparison with our data as accurate as possible. Indeed, most data has been obtained from droplets sliding on a tilted plane with a tilt angle between 40 and 50° (angle in our experiments has been 40°). The droplet volume $V_D$ and sliding path given on the right y axis are in the same or a similar range allowing a comparison with our tests. Charges on pristine are among the lower observed charge quantities likely due to the reduced surface contact on the superhydropbobic leaves. However, the treated leaves with smoother wax layer (static contact angle = 105°), are charging even better than most fluorinated surfaces. Only the experiments by Hinduja et al. \cite{hinduja_slide_2024} slightly larger charges than the best-performing leaf, but those measurements have been done by pulling an elliptical droplet of 4-times the volume as used in our experiments over the surface so that charges are likely higher by the increased surface contact. This renders our best-performing leaf the highest charging surface among those where droplets simply slide on the surface. The data in the figure are from references \cite{Li_2022, Xu_TriboelectricWetting_2022, stetten_slide_2019,hinduja_slide_2024,Chen_irriversible_2025,zhou_deposition_2025}}
  \label{ref_figsX}
\end{figure}

\pagebreak
\subsection{Recovery of initial droplet charge when the leaf is unexposed to droplet stimulation}
\begin{figure}[h!]
  \centering
  \includegraphics[width=\textwidth]{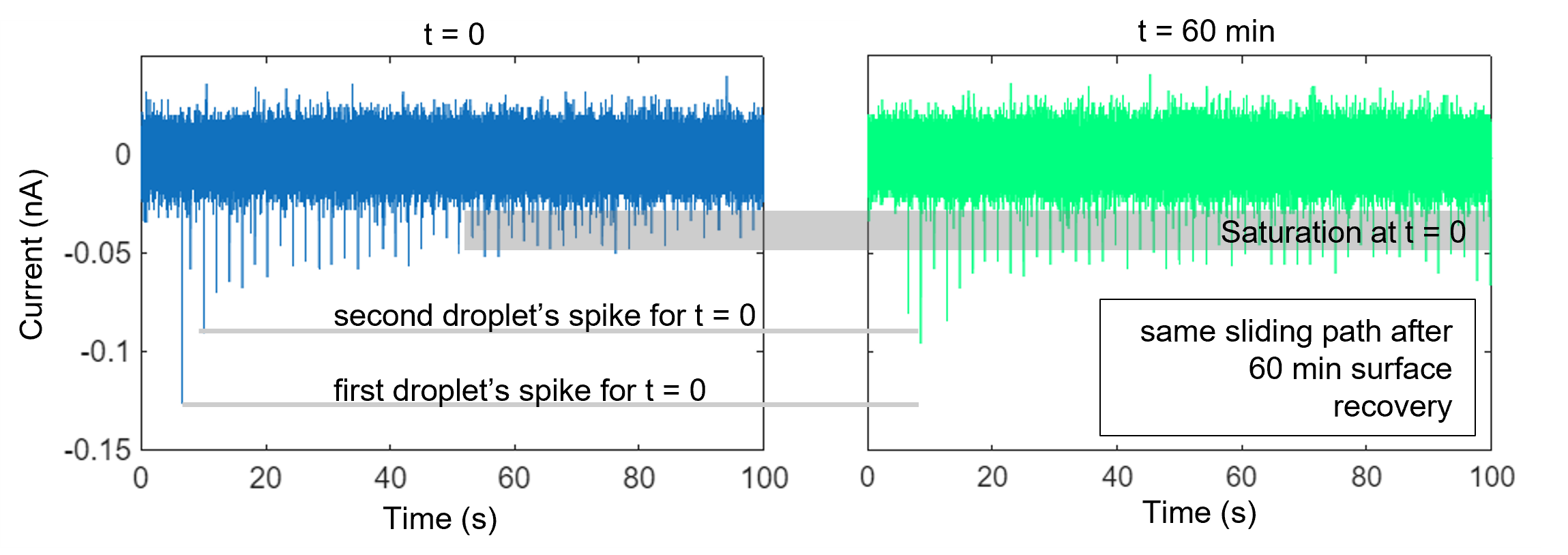} 
  \caption{Recovery of higher charges accumulated in first droplets after 60 min. The figure shows droplet discharge current measurements on a pristine leaf at $t=0$ and $t=60$ min showing that the signals initially reduced when sequential droplets hit the leaf (left panel, blue curve) and then saturate at a lower current. When the leaf is left for 60 min without droplet exposure, the initial droplets again acquire a higher current suggesting a recovery of sites for charge transfer.}
  \label{ref_figs8}
\end{figure}

\pagebreak
\subsection{Extended figure: Analysis of the resistive force acting on sequential droplets}

\begin{figure}[htbp!]
  \centering
  \includegraphics[width=0.75\textwidth]{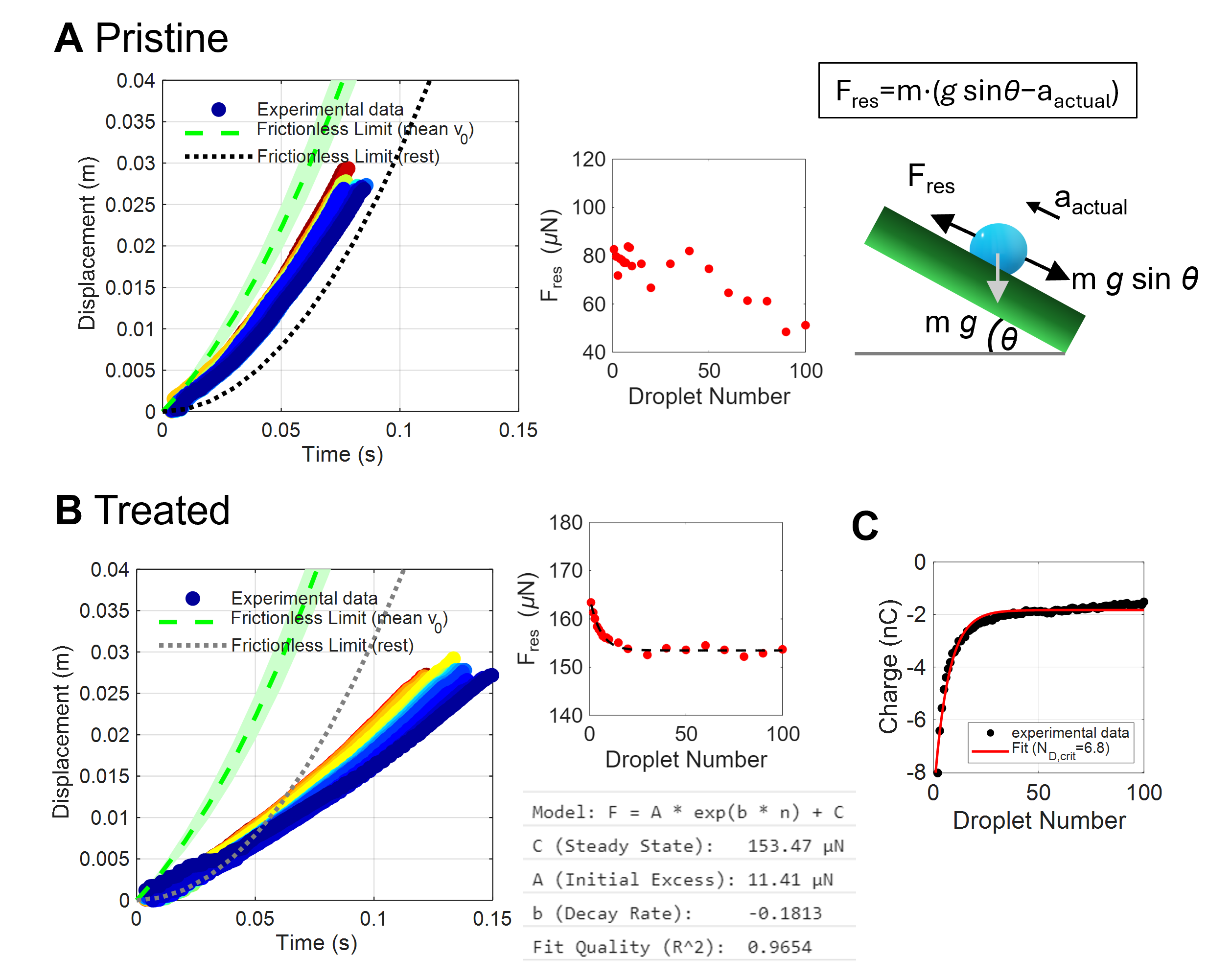} 
  \caption{Kinetic analysis of the resistive force acting on sequential droplets. a) Left plot: experimental data from pristine leaf surface of the displacement of a 30 $\mu$L droplet as function of time and droplet number (blue, D1 to red, D100 colored circles), calculated frictionless displacement given by the dotted black line for droplet at the leaf inclination angle $\theta = 40°$ when droplet slides from resting position $v_0 = 0$, and frictionless displacement assuming that the droplets have an initial speed $v_0$ of $0.29 \pm 0.06$ m/s caused during pumping of the droplet onto the leaf surface and estimated by extrapolating the first 25\% of the time-displacement curves of each droplet (dashed green line, shaded area represents the standard deviation of $v_0$ derived from all droplets). Central plot: instantaneous resistive force $F_{res}$ for each droplet calculated as the deviation from the ideal frictionless acceleration using the force balance depicted on the right of the plot and further described in the text. b) Left plot: Corresponding data after leaf surface treatment and variation of the epicuticular wax structure. Central plot: the instantaneous resistive force $F_{res}$ for each droplet on the treated surface. C. Droplet charge as function of droplet number of the treated leaf surfaces corresponding to the motion data given in B resulting in an $N_{D,crit}$ of 6.8. \label{ref_figs9}}
\end{figure}

\clearpage
\appendix

\setcounter{figure}{0}
\renewcommand{\thefigure}{S\arabic{figure}} 
\renewcommand{\theHequation}{\thesection.\arabic{equation}} 
\renewcommand{\theHfigure}{SM.\arabic{figure}} 

\renewcommand{\figurename}{Supporting Information Figure}
\setcounter{figure}{0} 
\renewcommand{\thefigure}{\arabic{figure}}

\begin{center}
    \LARGE \textbf{Supporting Information: Spontaneous epicuticular charging affects droplet dynamics on living leaves}
\end{center}

\section{Method: droplet tracking}
The experiments performed in this study yielded an extensive video dataset, comprising a total of over 300 videos. Each video recorded with high frame rate (1200 fps) is approximately 150 frames in length, carefully capturing the dynamic motion of droplets along the 40 mm slide path between two electrodes. To ensure the accuracy of subsequent velocity measurements, it was imperative to precisely track the center of mass of each droplet across every frame. Given the large video data and need to precisely track droplets, we employed DropTrack, a specialised computer vision software for droplet recognition and tracking, which leverages a combination of the YOLO and DeepSORT algorithms. The software was trained on a custom dataset created by randomly sampling 1,121 images from the total video collection. These images were then manually annotated with bounding boxes, meticulously outlining each droplet in the YOLO format using the Roboflow platform. This dataset was split in a 9:1 ratio between training and validation data.

\begin{figure}[htbp!]
  \centering
  \includegraphics[width=\textwidth]{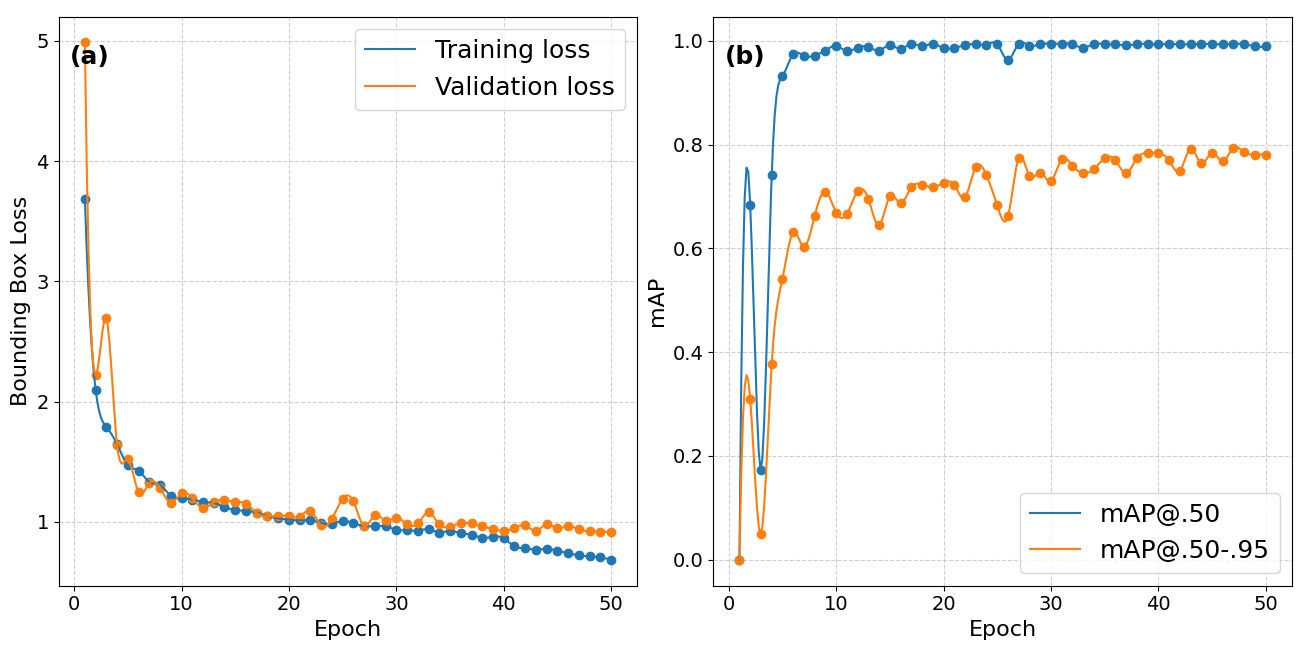}
  \caption{a) Evolution of bounding box loss during training. b) Training performance assessed by mean average precision (mAP) at Intersection over Union (IoU) thresholds 0.5 and 0.5 to 0.95. }
  \label{ref_figs1}
\end{figure}

Once the training dataset was assembled, a YOLO-based convolutional neural network was trained to detect droplets in the image sequences. Model training was performed using the implementation provided by Ultralytics, with code and documentation publicly available through their repository. \cite{yolo11_ultralytics}. Supporting Information Figure \ref{ref_figs1} illustrates the model's training progression. The plot displays the reduction in bounding box loss, which directly corresponds to the diminishing error in bounding box coordinate prediction, and the concurrent increase in mean average precision (mAP), a key metric for evaluating the overall quality and performance of the trained model. In this work, we deployed a YOLOv11l model in conjunction with the DeepSORT algorithm for our analysis. When applied to our custom dataset, the model's performance plateaued at a mean average precision (mAP) of 0.98 for an Intersection over Union (IoU) threshold of 0.5, and 0.78 for the more rigorous range of IoU thresholds from 0.5-0.95. These values stand higher to the mAP@0.5-0.95 scores of around 0.54 typically reported for general object recognition tasks \cite{yolo11_ultralytics}. The high mAP values achieved here, therefore, underscore the model's strong ability to precisely recognise droplets within the images. The pre-trained weight for YOLOv11 and the custom data as mentioned above can be found here. 

The trained DropTrack software takes a video as input and outputs a data file containing a unique identifier for each detected droplet, enabling its motion to be tracked, along with the coordinates of its bounding box in every frame. From this data, the droplet's center of mass was approximated as the geometric center of its bounding box. Consequently, the droplet's velocity was directly derived from tracking the speed of this central point. This same data also provided a straightforward method for measuring the droplet's width.

\section{Kinetic analysis of the resistive force acting on sequential droplets}
To further quantify the dynamic wetting behavior and the effect of the variation in the epicuticular wax layer plasticity on droplets sliding on the leaves, the overall resistive forces acting on the droplet by comparing the experimental trajectory against a theoretical frictionless slide was estimated as follows.

\subsection{Force Balance Model}
The motion of a droplet of mass $m$ sliding down a surface inclined at an angle $\theta$ is governed by the interplay between the gravitational force and the total resistive force $F_{res}$ (a sum of multiple coexisting forces such as contact line friction, viscous dissipation, and electrostatic forces). By Newton's Second Law applied parallel to the inclined leaf surface, the force balance is given by:

\begin{equation}
    m a_{actual} = m g \sin(\theta) - F_{res}
    \label{eq:force_balance}
\end{equation}

where $g$ is the gravitational acceleration and $a_{actual}$ is the observed acceleration of the droplet.

\subsection{Experimental Acceleration and Polynomial Fitting}
To determine $a_{actual}$, the time-displacement data $(t, d)$ extracted from the high-speed video analysis were fitted to a second-order polynomial, corresponding to the kinematic equation for constant acceleration:

\begin{equation}
    d(t) = p_1 t^2 + p_2 t + p_3 \quad \equiv \quad \frac{1}{2} a_{actual} t^2 + v_0 t + d_0
\end{equation}

The acceleration is derived directly from the quadratic coefficient $p_1$ of the fit:

\begin{equation}
    a_{actual} = 2 \cdot p_1
\end{equation}

\subsection{Calculation of Resistive Force}
\label{calc_resistive_forces}
Rearranging Eq. (\ref{eq:force_balance}), the instantaneous resistive force for each droplet was calculated as the deviation from the ideal frictionless acceleration ($a_{ideal} = g \sin\theta$):

\begin{equation}
    F_{res} = m (g \sin\theta - a_{actual})
    \label{eq:Fres}
\end{equation}

Extended Data Figure \ref{ref_figs9} show the results for pristine (upper panel A) and treated leaves (lower panel B). The graph on the left shows the data points for D1 to D100 (blue to red colored circles) that have been fitted to extract $a_{actual}$, the correlating $F_{res}$ is displayed as function of drop number in the smaller graph on the right. For comparison, the theoretical frictionless displacement (dotted black line) was calculated for an inclination of $\theta=40°$ assuming a resting start ($v_0=0$). However, the experimental droplets exhibited velocities exceeding this frictionless limit, indicating a significant certain velocity imparted by the peristaltic pump. This initial speed ($v_0$) was quantified by extrapolating the first 25\% of the time-displacement curves on pristine superhydrophobic leaves, where frictional losses are minimal. The resulting $v_0$ was determined to be $0.29 \pm 0.06$ m/s (dashed green line, the shaded area represents the standard deviation).

By deriving the resistive force $F_{res}$ from the acceleration, the influence of the initial velocity was eliminated. The analysis reveals that for both pristine and treated leaves, $F_{res}$ decays as a function of droplet number. Notably, the initial resistive force is highest for the first droplet, measuring 83 $\mu$N on pristine leaves and approximately double that value (164 $\mu$N) on treated leaves. These values decline to about 50 $\mu$N on pristine and 150 $\mu$N on treated leaves, respectively for D100 indicating a significant decrease in the resistive forces among the first 100 droplets sliding on the leaves. Interestingly, the surface treatment and resulting wax layer variation not only strongly increase the resistive forces but also introduce a droplet-number dependent exponential decay in $F_{res}$ reaching saturation after about 20-30 droplets (the evolution of $F_{res}$ as a function of droplet number $n$ was modeled using an exponential decay function: $F_{res}(n) = A \cdot e^{b \cdot n} + C$ where $C$ represents the value of $F_{res}$ on the saturated surface (e.g., $n>100$), and $A$ represents the difference of the saturation state to  $F_{res}$ at D1, b is the decay constant). The derived parameters are given in the Extended Data Figure \ref{ref_figs9}B. Interestingly, comparing the decay constant $b$ ($=0.18$) obtained from the kinetic video analysis with the charge saturation rate $1/N_{D, \text{crit}}$ ($1/6.8 \approx 0.15$) from simultaneous measurements reveals very similar kinetics. This suggests that surface charging of the leaf significantly correlates with the resistive force, particularly for the highly charged initial droplets. Assuming that the difference between droplet 1 $F_{res, D1}$ and $F_{res, Saturation}$ at saturation is due to an electrostatic force, the resulting force is approximately 11 $\mu$N. This is in a comparable range (albeit 2 to 4 times lower) to electrostatic resistive forces measured on artificial fluorinated surfaces (20-80 $\mu$N) \cite{Li_2022}  under similar experimental conditions. 
\begin{table}[h]
    \centering
    \caption{Description of variables used in the kinetic analysis.}
    \label{tab:variables}
    \begin{tabular}{l l l}
        \hline
        \textbf{Symbol} & \textbf{Description} & \textbf{Unit} \\
        \hline
        $m$ & droplet mass& kg \\
        $\theta$ & Inclination angle of the surface & degrees ($^\circ$) \\
        $g$ & Gravitational acceleration ($9.81$) & m/s$^2$ \\
        $a_{ideal}$ & Theoretical acceleration on a frictionless surface & m/s$^2$ \\
        $a_{actual}$ & Experimentally measured acceleration & m/s$^2$ \\
        $F_{res}$ & Total resistive forces& N \\
        $p_1$ & Quadratic coefficient from polynomial fit & m/s$^2$ \\
        $n$ & Droplet sequence number & - \\
        \hline
    \end{tabular}
\end{table}

\section{Supporting videos}
In this study, we recorded and analysed more than 300 videos depicting the motion of droplets across a pair of electrodes. The included videos showcase both the original footage and the corresponding droplet tracking data. \\

\noindent Video 1: An original video showing a droplet travelling between two electrodes on a pristine leaf. \\
Video 2: The visual depiction of the droplet tracking process on a pristine leaf.  \\
Video 3: Comparison between droplets travelling on pristine and treated leaf surfaces. The videos show motion of droplet 1 on the same leaf and sliding zone before (pristine) and after treatment (treated).  \\
Video 4: An original video showing droplets travelling between two electrodes on a treated leaf with the visual depiction of the droplet tracking process.\\
Video 5: Comparison of the difference in motion of droplet 1 and droplet 100 on the pristine leaf. 
Video 6: Comparison of the difference in motion of droplet 1 and droplet 100 on the treated leaf.

\section{Data availibility}
The pre-trained weight for YOLOv11, the custom training data, and supporting videos as mentioned above as well as the droplet charge measurements can be found here 10.5281/zenodo.17827801.


\end{document}